\documentclass[conference]{IEEEtran}
\IEEEoverridecommandlockouts
\usepackage{cite}
\usepackage{amsmath,amssymb,amsfonts}
\usepackage{algorithmic}
\usepackage{graphicx}
\usepackage{textcomp}
\usepackage{xcolor}
\usepackage{makecell}
\usepackage{hyperref}
\usepackage{pifont}
\usepackage{multirow}

\usepackage{caption}
\usepackage{enumitem}
\usepackage{orcidlink}

\def\BibTeX{{\rm B\kern-.05em{\sc i\kern-.025em b}\kern-.08em
    T\kern-.1667em\lower.7ex\hbox{E}\kern-.125emX}}
\begin{document}

\title{
Optimizing Semiconductor Device Simulations through Low-Precision Arithmetic
}

\author{
\IEEEauthorblockN{
Alexander Maeder\,\orcidlink{0009-0003-4420-5593},
Denghui Lu\,\orcidlink{0000-0003-0977-3635},
Nicolas Vetsch\,\orcidlink{0000-0002-0818-8461},
Vincent Maillou\,\orcidlink{0000-0003-4861-3298},
Anders Winka\,\orcidlink{0009-0001-9256-1256},\\
Jiang Cao\,\orcidlink{0000-0002-7646-8318},
Mauro Dossena\,\orcidlink{0000-0001-5358-2452},
Alexandros Nikolaos Ziogas\,\orcidlink{0000-0002-4328-9751},
Mathieu Luisier\,\orcidlink{0000-0002-2212-7972}
}
\IEEEauthorblockA{\textit{D-ITET, ETH Zurich}, Zurich, Switzerland \\
\{almaeder, denghuilu, vetschn, vmaillou, awinka, jcao, mdossena, alziogas, mluisier\}@iis.ee.ethz.ch}
}



\maketitle

\begin{abstract}
Architectural changes in GPUs, especially the promotion of low-precision computational units, pose significant challenges to traditional, FP64-based high-performance computing (HPC) applications, while also presenting opportunities.
Adopting reduced-precision data formats is a promising avenue to exploit the increased throughput capabilities. However, straightforward data conversions may lead to degraded accuracy or even erroneous results. For a given application, only an in-depth analysis of its numerical stability can reveal the potential of low-precision arithmetic. In this work, we consider the open-source \textit{quatrex} package, a quantum transport solver capable of breaking the sustained FP64 Eflop/s barrier, to illustrate trade-offs between accuracy losses and computational speed-ups when moving from high- to low-precision formats.
We use three representative benchmark structures to explore the application's numerical properties.
Applying the gained insights to a larger, more realistic system, we achieve up to 51\% higher throughput while maintaining accurate results, on 40\% fewer HPC resources than the FP64 reference.

\end{abstract}


\vspace{-.5em}
\section{Introduction}
\label{sec:intro}

Traditional high-performance computing (HPC) applications largely rely on double-precision floating-point (FP64) arithmetic to ensure sufficiently high accuracy.
This is the primary reason for highlighting and tracking over time the FP64 performance of supercomputers~\cite{top500}.
At the same time, rapid developments in machine learning (ML) have driven significant strides in research toward lower- and mixed-precision arithmetic.
Combined with the deployment of large language models (LLMs), ML has significantly altered the dynamics of the hardware market, leading architecture designers to increasingly focus on accelerating lower-precision arithmetic operations, potentially at the expense of high-precision silicon.
For example, the NVIDIA Blackwell-based GB200 chips~\cite{nvidia_gb200_ds} have lower FP64 performance than their predecessors, the Hopper-based GH200~\cite{nvidia_gh200_ds}, although they provide two times larger lower-precision throughput.
While these recent developments may cast doubts on the future of hardware-accelerated high-precision arithmetic, they also present unique opportunities for HPC applications to take advantage of this paradigm shift.
Indeed, leveraging low-precision arithmetic might help reduce both execution times and memory footprints.

As a representative application that utilizes FP64 arithmetic but could benefit from low-precision arithmetic, we consider the open-source \textit{quatrex} code~\cite{qtx-2024}, a so-called quantum transport (QT) solver dedicated to the simulation of electronic currents through nanoscale semiconductor devices. Accurately modeling the ``current vs. voltage'' (``I-V'') characteristics of, e.g., industry-relevant silicon nano-ribbon field-effect transistors (NRFET)~\cite{intel-nrfet}, requires the inclusion of quantum mechanical (quantization and tunneling) and scattering (carrier-carrier or carrier-phonon) effects as well as the treatment of their geometry with atomic resolution. Traditional technology computer-aided design (TCAD) tools based on (semi-)classical physics fail at capturing these critical properties. They should, therefore, be gradually replaced by more advanced packages such as \textit{quatrex} that can validate novel materials and nano-device concepts \textit{in-silico}. Notably, \textit{quatrex} allowed for the first QT simulation of realistic NRFETs, with dimensions comparable to experimental structures, in the presence of carrier-carrier interactions, reaching a sustained FP64 performance of 1.146 Eflop/s and 342 Pflop/s on the Frontier (AMD Instinct MI250X GPUs and AMD EPYC 64-core Trento CPUs) and Alps (GH200 superchips) supercomputers, respectively~\cite{qtx-2025}.

Presently, the overall cost of producing the ``I-V'' curve of an NRFET at a quantum mechanical level, with scattering, is immense, despite the high computational performance and parallel efficiency of \textit{quatrex}. To simulate the electronic current corresponding to a single bias (V) point, the computational complexity is equal to $\mathcal{O}\left(N_{iter}N_{E}N_{A}^3\right)$ on $\mathcal{O}\left(N_{E}N_{A}^2\right)$ space, where $N_{A}$ is the total number of atoms composing the simulated device, $N_{E}$ is the number of different energies that electrons can adopt, and $N_{iter}$ refers to the number of iterations the physical models need to converge.
For the aforementioned state-of-the-art NRFETs made of $N_A=42,240$ atoms and $N_E=18,800$ energy points, the computational workload reaches 48.252 exaflop per iteration, while the memory footprint is on the order of a few petabytes~\cite{qtx-2025}. Achieving such a high energy resolution could only be realized at the full scale of Frontier, currently the second-largest supercomputer in the world.
These results, although promising, demonstrate that advanced TCAD tools might only become integral parts of routine semiconductor design processes if the underlying computational complexity is substantially decreased.

Lower-precision arithmetic appears as a practical way to relax these very high computational and memory requirements.
Storing the main quantities in lower-precision data types can significantly reduce the space complexity, enabling QT simulations on memory-constrained systems.
Furthermore, mixed-precision computations can further accelerate the execution of pivotal compute-bound kernels.
On the other hand, applying such techniques may lead to dramatic losses of numerical accuracy, potentially affecting the final results.
Certain operations, for example, matrix inversions or eigenvalue problems, may be highly sensitive to perturbations in their inputs, especially when these are ill-conditioned.
The iterative nature of QT solvers may further amplify such errors if they propagate from one iteration to the next.
Hence, before applying lower-precision arithmetic to TCAD tools such as \textit{quatrex}, it is imperative to study their numerical properties in great detail and devise a suitable strategy from this analysis.

To assess the potential of lower-precision arithmetic in QT simulations, we conduct in this work an empirical yet comprehensive study of the numerical stability of the \textit{quatrex} package~\cite{qtx-2024,qtx-2025}.
Its physical models articulate themselves around the non-equilibrium Green's function (NEGF) formalism~\cite{datta_1995} combined with carrier-carrier interactions expressed in the well-established GW approximation~\cite{Hybertsen1986}.
The equations pertaining to these carrier-carrier interactions lead to scattering self-energies~\cite{Thygesen2008} that must be treated self-consistently (sc) with the NEGF system.
All required input quantities (Hamiltonian, overlap, and Coulomb matrices as well as atomic geometries) are obtained from density functional theory (DFT) calculations. DFT is a powerful and widely used \textit{ab initio} method to determine the structural, electronic, and vibrational properties of solids~\cite{dft}.
In the following, we will refer to this combined QT model implemented in \textit{quatrex} as DFT+NEGF+scGW.





\begin{figure}[h]
\vspace{-2.5em}
  \centering

  \includegraphics[width=0.48\textwidth]{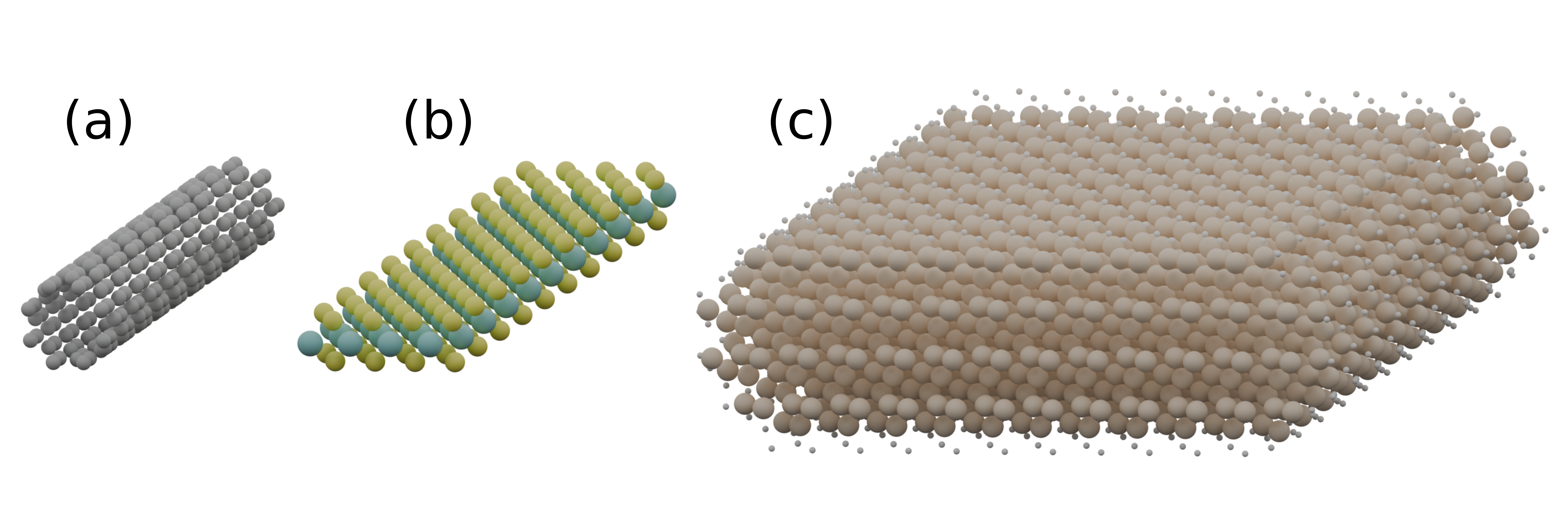}

  \vspace{-0.5em}

\small
\centering
  \begin{tabular}{ l l c c c || c}
    Property & Description & CNT & MoS$_2$ & \mbox{NR-S} & \mbox{NR-L}\\ 
    \hline
    $N_{A}$ &   \makecell[l]{\footnotesize Number of\\\footnotesize atoms}\small & 768 & 288 & 1,584 & 10,560\\
    \hline
    $N_{orb}$ & \makecell[l]{\footnotesize Number of\\\footnotesize atomic orbitals}\small  & 768 & 1,056 & 5,112 & 34,080 \\
    \hline
  \end{tabular}

  \caption{\normalsize Schematic view of test structures and their parameters: (a) carbon nanotube (CNT), (b) molybdenum disulfide (MoS$_2$), and (c) silicon nanoribbon (\mbox{NR-S} / \mbox{NR-L}).
  }
  \label{fig:devices}
  \vspace{-1em}
\end{figure}

To gauge whether, and if so how, the precision requirements of this model can be reduced, we consider three nano-transistor designs that are currently of high relevance to the semiconductor industry: (a) a carbon nanotube (CNT)~\cite{cnt_vlsi_2023}, (b) a single layer of molybdenum disulfide (MoS$_2$)~\cite{mos2_iedm_2025}, and (c) an NRFET channel~\cite{intel-nrfet}. Their atomistic structures and the simulation parameters employed in our analysis are shown in Fig.~\ref{fig:devices}.
To enable execution of the full solver with a relatively limited amount of computational resources (approximately 40 to 512 NVIDIA GH200 superchips), the device dimensions are kept intentionally small. We consider structures comprising between 288 (MoS$_2$) and 10,560 (\mbox{NR-L}) atoms, with a total number of orbitals $N_{orb}$ between 768 and 34,080.

Using these representative devices as testbeds, we investigate the effect of different lower-precision storage and computational schemes on the accuracy of the simulator's output. In our study, we aim to address the following questions:
(i) Does the simulator still reach convergence when leveraging lower-precision arithmetic?
(ii) If it does, is the rate at which it converges affected?
(iii) What is the relative error of the simulation results with respect to a baseline FP64 calculation?
(iv) What are the benefits of lower-precision arithmetic in terms of computational throughput?
We summarize our contributions:
\begin{itemize}
    \item We examine the impact of lower-precision arithmetic on the accuracy of the computed electronic currents for three representative nano-transistor configurations (CNT, MoS$_2$, \mbox{NR-S}) with up to 1,584 atoms.
    \item We accelerate computations via floating-point emulation in matrix-matrix products (Ozaki scheme II~\cite{ozaki-2}), guided by insights gained from our accuracy studies.
    \item We apply our lower-precision QT scheme to a larger NRFET (\mbox{NR-L}) comprising 10,560 atoms.
    A reduction of up to 40\% in required compute resources,
    and a 51\% throughput compared to the FP64 reference are achieved.
\end{itemize}

The rest of this paper is organized as follows:
Section~\ref{sec:back} presents the basic formulation of the DFT+NEGF+scGW model, along with its intermediate quantities and algorithmic motifs. In Section~\ref{sec:memory}, we probe how lower-precision \textit{storage} affects convergence and accuracy of QT simulations. A similar analysis is performed in Section~\ref{sec:compute} for lower-precision \textit{computation} schemes.
Based on these findings, we implement a novel mixed-precision scheme in \textit{quatrex} and evaluate it in Sections~\ref{sec:evaluation} and \ref{sec:performance} for a larger, experimentally-relevant nano-transistor structure.
We close by discussing related work in Section~\ref{sec:related}, addressing limitations in Section~\ref{sec:limitations}, and providing an outlook in Section~\ref{sec:conclusions}.

\vspace{-.5em}
\section{Background}
\label{sec:back}

\subsection{The DFT+NEGF+scGW approach}
\label{sec:back-dft+ngf}

In QT frameworks, open quantum systems are driven out of equilibrium, for example, by applying an external voltage. This perturbation gives rise to a flow of particles, e.g., electrons, from one electrode (contact) to another one. In other words, the voltage acts as a driving force for the electronic current $I_d$.
The state-of-the-art method to model electron transport at a quantum mechanical level and from first-principles is DFT+NEGF~\cite{stokbro}, its main advantage being that ballistic conditions as well as interactions with crystal vibrations~\cite{frederiksen}, defects~\cite{Dossena2025}, or other charge carriers~\cite{Thygesen2008} can be treated in a unified manner. Carrier-carrier interactions, in particular, are crucial when modeling ultra-scaled transistors, where high populations of interacting electrons are confined to small volumes. These interactions can affect device functionality in non-trivial ways, for instance, by significantly decreasing the magnitude of the electronic current~\cite{deuschle2025electron}. To accurately design future semiconductor devices, it is thus essential to account for carrier-carrier scattering through the self-consistent GW (scGW) approximation, as implemented in \textit{quatrex}~\cite{qtx-2025}.

The central quantities in the DFT+NEGF+scGW approach are the so-called lesser ($\mathbf{G}^{<}$) and greater ($\mathbf{G}^{>}$) Green's functions. They are key to obtaining the physical observables of interest, e.g., the electronic current, the electron/hole population, or the electrostatic potential~\cite{deuschle2025electron}.
As already hinted in the previous section, the evaluation of these Green's functions requires self-consistency, as the underlying physics lead to a self-referential, non-linear equation of the form ${\mathbf{G}=\mathcal{F}(\mathbf{G})}$.
In practice, the evaluation of $\mathbf{G}$ follows an iterative approach, where every iteration involves four distinct phases, each pertaining to a different intermediate quantity
\begin{equation}
    \mathbf{G}\left[ \mathbf{\Sigma} \right] \rightarrow
    \mathbf{P}\left[ \mathbf{G} \right] \rightarrow
    \mathbf{W}\left[ \mathbf{P} \right] \rightarrow
    \mathbf{\Sigma}\left[ \mathbf{G}, \mathbf{W} \right].
\label{eq:self_consistent}
\end{equation}
Here $\mathbf{P}$ is the polarization, $\mathbf{W}$ denotes the screened Coulomb interaction, and $\mathbf{\Sigma}$ refers to the carrier-carrier scattering self-energy. 
The inter-dependencies between $\mathbf{G}$, $\mathbf{P}$, $\mathbf{W}$, and $\mathbf{\Sigma}$ are expressed with square brackets.
All quantities can be of lesser ($<$), greater ($>$), retarded ($R$), and advanced ($A$) type~\cite{Kadanoff2018}, which is usually denoted as superscript (e.g. $\mathbf{\Sigma}^{\lessgtr}$) but is omitted in Eq.~(\ref{eq:self_consistent}) for brevity.
Furthermore, all quantities are discretized on a grid of energies $E$. For each energy point, they take the form of a sparse matrix of size $N_{orb}\times N_{orb}$, where $N_{orb}$ is the total number of atomic orbitals in the system under test. In QT simulations, these quantities can be stored as $N_E\times N_{orb}\times N_{orb}$ distributed tensors. Since they all exhibit either skew-Hermitian or general complex symmetry, only their upper half needs to be stored explicitly. Furthermore, since the matrix elements describe interactions between atomic orbitals, which generally attenuate with increasing distance between them, all quantities exhibit a quasi-banded sparsity pattern that can be tiled into a block-tri-diagonal (BT) form.

Each of the four phases in a self-consistency cycle falls into one of two categories: a selected solution of a quadratic matrix equation or an energy convolution. The Green's functions $\mathbf{G}$ and the screened Coulomb interactions $\mathbf{W}$ are given by
\begin{equation}
\mathbf{M}(E)\mathbf{X}^{\lessgtr}(E)\mathbf{M}(E)^{\dagger} = \mathbf{B}^{\lessgtr}(E),
\label{eq:system-solve}
\end{equation}
which must be solved independently for each energy $E$. The solution of Eq.~(\ref{eq:system-solve}) can be accelerated as only selected entries of $\mathbf{X}^{\lessgtr}(E)$ are needed, namely those corresponding to the BT sparsity pattern of $\mathbf{M}(E)$ and $\mathbf{B}^{\lessgtr}(E)$~\cite{maillou2026parallelquadraticselectedinversion}.
The entries of the scattering self-energy $\mathbf{\Sigma}$ and polarization $\mathbf{P}$ result from element-wise energy convolutions
\begin{equation}
\mathbf{A}_{ij}^{R,\lessgtr}(E) \propto \int dE'\, \mathbf{Y}_{ij}^{\lessgtr}(E-E') \mathbf{Z}_{ij}^{\gtrless}(E'),
\label{eq:interaction}
\end{equation}
which can be evaluated efficiently via fast Fourier transforms (FFT)~\cite{Thygesen2008}.
Typically, energy resolutions of 5~meV are sufficient to obtain accurate simulation results, but as the energy range extends over tens of eV, the total number of energy points is in the order $N_E\propto10,000$.

In summary, the computation of the $\mathbf{G}$, $\mathbf{P}$, $\mathbf{W}$, and $\mathbf{\Sigma}$ tensors forms a chain of quadratic systems and convolutions, that are embarrassingly parallel over energies $E$ and over orbital indices $ij$, respectively, with data being communicated between each of the four phases via Alltoall collectives~\cite{qtx-2025}. The specific form of each equation is given in Table~\ref{tab:quantities}, where the quantities are grouped by computational motif.

\begin{table}[!t]
\caption{
Definition of quantities in Eqs.~(\ref{eq:system-solve}) and (\ref{eq:interaction}). The Hamiltonian $\mathbf{H}_{DFT}$, the overlap matrix $\mathbf{S}_{DFT}$, and the Coulomb matrix ($\mathbf{V}$) are obtained via DFT calculations.
}
\small
\def\arraystretch{1.3}
    \centering
    \begin{tabular}{c||c|c}
         & $\mathbf{M}(E)$ & $\mathbf{B}^{\lessgtr}(E)$\\
        \hline
        \hline
         \makecell[c]{Green's Function\\$\mathbf{X}^{\lessgtr} \equiv \mathbf{G}^{\lessgtr}$}& \makecell[c]{$E\mathbf{S}_{DFT} - \mathbf{H}_{DFT}$\\$- \mathbf{\Sigma}^R(E)$} & $\mathbf{\Sigma}^{\lessgtr}(E)$\\ 
        \hline
        \makecell[c]{Screened Coulomb\\$\mathbf{X}^{\lessgtr} \equiv \mathbf{W}^{\lessgtr}$} & $\mathbf{I} -\mathbf{V}\mathbf{P}^{R}(E)$ & $\mathbf{V}\mathbf{P}^{\lessgtr}(E)\mathbf{V}^{\dagger}$\\ 
        \hline
        \multicolumn{3}{c}{} \\
         & $\mathbf{Y}^{\lessgtr}(E)$ & $\mathbf{Z}^{\lessgtr}(E)$\\
        \hline
        \hline
         \makecell[c]{Polarization\\$\mathbf{A}^{R,\lessgtr} \equiv \mathbf{P}^{R,\lessgtr}$} & $\mathbf{G}^{\lessgtr}(E)$ & $\mathbf{G}^{\lessgtr}(E)$\\        
        \hline
        \makecell[c]{GW Self-Energy\\$\mathbf{A}^{R,\lessgtr} \equiv \mathbf{\Sigma}^{R,\lessgtr}$} & $\mathbf{G}^{\lessgtr}(E)$ & $\mathbf{W}^{\lessgtr}(E)$\\ 
        \hline
       \end{tabular}
    \label{tab:quantities}
    \vspace{-1.5em}
\end{table}


\subsection{Accuracy of QT device simulations}
\label{sec:accuracy}

The accuracy of QT device simulations is linked to the electronic current $I_d$ flowing across a given device structure: The $I_d$ values extracted at different locations along the transport direction should not vary with respect to one another by more than a pre-defined criterion, $I_{d,conv}$, set at the beginning of each simulation. This condition, known as ``current conservation'', is equivalent to Kirchhoff's first circuit law, which states that the current entering any junction is equal to the current leaving this junction \cite{kirchhoff}. Once it is satisfied, the $\mathbf{G}\rightarrow\mathbf{P}\rightarrow\mathbf{W}\rightarrow\mathbf{\Sigma}$ cycle stops. From a numerical perspective, since $I_d$ can be directly derived from $\mathbf{G}$, convergence of the electronic current implies that the variation of entries in the Green's function tensors between two consecutive iterations $n-1$ and $n$ does not exceed a tolerance~$\epsilon$,
\begin{equation}
\left\lVert \mathbf{G}_{n} - \mathbf{G}_{n-1} \right\rVert < \epsilon.
\end{equation}
In practice, it is typically assumed that $I_{d,conv}\propto10^{-2}$ is a satisfactory convergence criterion \cite{deuschle2025electron}. 
Such a value is primarily justified by the fact that current conservation of this magnitude is a clear indication that all intermediate variables in Eq.~(\ref{eq:self_consistent}) have become stationary. 
Hence, any implementation of the DFT+NEGF+scGW method, specifically low-precision ones, should not induce errors in the same order of magnitude as this tolerance. In all our analyses of low-precision schemes, we will ensure that no relative errors above $10^{-3}$ are introduced.

\subsection{Floating-point arithmetic}
\label{sec:back-fp}

Floating-point representation allows for a wide dynamic range of values. It partitions a fixed number of bits into three components: a sign bit (s), a number of bits encoding the exponent (E), and bits representing the significand (p).
Following the IEEE-754 standard~\cite{ieee-754}, the value ($v$) of a floating-point number is expressed as 
\begin{equation}
v = (-1)^\text{s} \times (1.\text{p}) \times 2^{\text{E} - \text{B}},
\label{eq:floating_point}
\end{equation}
where B is a fixed exponent bias.
The format's precision is determined by the number of bits in the significand, while the dynamic range depends on the number of bits allocated to the exponent.
While 16-, 32-, 64-, 128-, and 256-bit formats are defined in the IEEE-754 standard, the rise of machine learning has led hardware vendors to introduce specialized formats like TensorFloat-32 (TF32) and bfloat16 (BF16) \cite{Fasi2021}.
These formats are designed to exploit the prevalence of lower-precision arithmetic in machine learning, while delivering significantly higher performance via specialized hardware, e.g., tensor cores.
Architectures such as the NVIDIA GH200 provide performance gains of $\sim7\times$ to $\sim14\times$ when using these formats instead of traditional FP64 \cite{nvidia_gh200_ds}. Table~\ref{tab:floating_point} provides an overview of the number of bits allocated to the constituent partitions in several standard and specialized floating-point formats as well as the corresponding machine epsilon. 

\begin{table}[!t]
\caption{Bit counts in IEEE-754 and specialized formats.}
\small
\def\arraystretch{1.3}
    \centering
    \begin{tabular}{c|c|c|c|c}
        Format & Total & Exponent & Significand &  \makecell[c]{Rounding\\ Machine Epsilon} \\
        \hline
        FP64 & 64 & 11 & 52 & $\sim1.11 \times 10^{-16}$ \\
        FP32 & 32 & 8 & 23 & $\sim5.96\times 10^{-8}$  \\
        FP16 & 16 & 5 & 10 & $\sim4.88\times 10^{-4}$  \\
        TF32 & 19 & 8 & 10 & $\sim4.88\times 10^{-4}$   \\
        BF16 & 16 & 8 & 7 & $\sim3.9\times 10^{-3}$ \\
        \hline
       \end{tabular}
    \label{tab:floating_point}
    \vspace{-1.5em}
\end{table}

\subsection{Floating-point emulation}
\label{sec:back-emu}

Instead of directly performing certain numerical operations using low-precision arithmetic, which can cause accuracy losses, applications may take advantage of lower-precision tensor cores through floating-point emulation schemes.
For example, the Ozaki schemes~I and II~\cite{ozaki-1,ozaki-2} enable high-precision matrix multiplication by leveraging lower-precision arithmetic.
These algorithms decompose the operands into many lower-precision slices and compute partial products of those slices, which are recombined into the final high-precision output.
A higher number of slices increases both the emulation precision and the computational cost.
In the context of Ozaki scheme~II, the slices are typically referred to as moduli, but, for simplicity, we will also denote them as slices.
State-of-the-art implementations of the Ozaki schemes~\cite{ootomo2024dgemm, uchino2025performance, ozaki-2, schwarz2026guaranteed} rely on INT8 tensor cores that, for example, in NVIDIA's GH200, achieve throughput of up to 1,979\;Top/s compared to the 67\;Tflop/s of FP64 tensor core performance on the same processor.
Therefore, even if these emulation algorithms involve (many) more operations than standard matrix multiplication routines, they still complete within shorter runtimes, especially when double-precision accuracy is not required.




\section{Floating-point precision for data storage}
\label{sec:memory}

DFT+NEGF+scGW implementations typically use a complex double-precision floating-point format (COMPLEX128) to store all intermediate ($\mathbf{G}$, $\mathbf{P}$, $\mathbf{W}$, $\mathbf{\Sigma}$, $\mathbf{M}$, $\mathbf{B}$) and input ($\mathbf{H}_{DFT}$, $\mathbf{V}_{DFT}$, $\mathbf{S}_{DFT}$) data. We will refer to the combination of all these quantities as DFT+NEGF+scGW data.
These quantities are very large in size ($N_E\times N_{orb}\times N_{orb}$ or $N_{orb}\times N_{orb}$).
Moreover, as they are interdependent, they have long, overlapping allocation lifetimes across each self-consistent iteration. They are, therefore, responsible for a massive memory footprint. 
Our first goal is to empirically determine the number of bits required to store these variables and to investigate how precision losses in data storage can affect simulation results. To that end, we consider the benchmark devices introduced in Section~\ref{sec:intro}: CNT, MoS$_2$, and \mbox{NR-S}.

\subsection{Methods}
\label{sec:memory-methods}

\begin{figure*}[t] 
  \centering
  \includegraphics[width=1\textwidth]{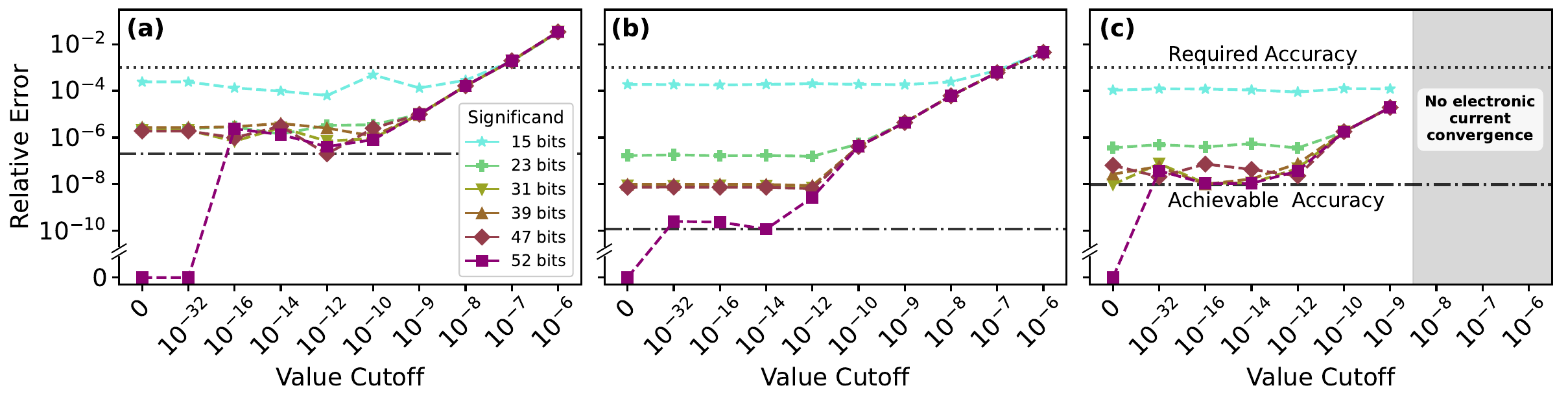}
  \caption{ 
  \normalsize
  Range and precision studies of DFT+NEGF+scGW for the (a) CNT, (b) MoS$_2$, and (c) \mbox{NR-S} devices from Fig.~\ref{fig:devices}. The lines with symbols indicate the relative error in the final electronic current ($I_d$), relative to the FP64 baseline, for various truncation schemes of the significand. The $x$-axis refers to the cutoff applied to the magnitude of the DFT+NEGF+scGW data.
  The horizontal black lines mark the lowest achievable error (`error floor') for each structure and the required accuracy for QT simulations.
  Note that 52 and 23 bits correspond to the standard FP64 and FP32 data types, respectively.
  }
  \label{fig:heatmaps}
  \vspace{-1.5em}
\end{figure*}

To restrict the dynamical range needed to represent all quantities, we have to first establish the range of floating-point values that the DFT+NEGF+scGW data can take.
We therefore simulate the previously mentioned three devices and record the maximum and minimum absolute values of all intermediate quantities, separately for the real and imaginary parts.
We find that in all cases the largest magnitude of both the real and complex parts is around $10-100$, with $10^3$ as an upper bound.
A clear lower bound cannot be identified, as arbitrarily small values, still representable in FP64, appear frequently in the data.
To instead define an ``effective'' lower bound, we introduce a cutoff, explicitly setting values with a magnitude below this cutoff to zero.
For the three devices, we find effective lower bounds around $10^{-29}$ (CNT), $10^{-39}$ (MoS$_2$), and $10^{-36}$ (\mbox{NR-S}), which, when enforced, do not alter the simulation results whatsoever: The computed electronic current remains bitwise identical to the FP64 reference.
Putting everything together, we experimentally determine a practical range for the absolute values of the DFT+NEGF+scGW data, namely $\left[10^{-39}, 10^3\right]$.
We note that this range approximately matches that of the FP32 format, namely an 8-bit exponent covering a range of normal numbers with magnitudes from $1.18\times10^{-38}$ to $3.4\times10^{38}$.

We further study the effect of lower-precision representation in two ways: (1) by progressively increasing the value cutoff, and (2) by truncating the least significant bits (LSBs) of the significand.
When truncating LSBs, we restrict ourselves to numbers of bits that, when combined with an 8-bit exponent and a sign bit, result in a custom byte-aligned floating-point format.
This allows us to adopt an efficient data compression method, which will be explained and showcased in Section~\ref{sec:evaluation}.
The considered significand lengths, in order of decreasing precision, are 47, 39, 31, 23, and 15.
Note that, at this stage, we modify only the precision in which the data is stored. All computations are still performed in FP64 to avoid conflating the effects of lower-precision data representations with precision loss in numerically sensitive algorithms.

\subsection{Storage precision analysis}
\label{sec:memory-studies}

Figure~\ref{fig:heatmaps} presents the relative error of the converged electronic current with respect to the FP64 baseline for the three devices under test.
The relative error is shown as a function of the value cutoff and for different numbers of significand bits.
The cutoff ranges from $0$ (no cutoff) to $10^{-6}$.
The number of bits in the significand is gradually decreased from 52 to 15, as explained in the previous section.

Although the relative errors tend to increase as the number of significand bits decreases, the DFT+NEGF+scGW method still reaches convergence. This happens even when truncating the significand to 15 bits only.
Nevertheless, we observe catastrophic loss of precision for the \mbox{NR-S} structure when imposing cutoff values greater than $10^{-9}$. In this regime, the self-consistent iterations no longer converge.

We also note that the smallest achievable relative error, i.e., the `error floor', is approximately $10^{-6}$ for the CNT, $10^{-10}$ for MoS$_2$, and $10^{-8}$ for \mbox{NR-S}.
These smallest relative errors place an effective upper bound on the simulations' achievable accuracy. They are indicated with a dash-dotted black line for each tested structure in Fig.~\ref{fig:heatmaps}.
Notably, the error floors are attained by a wide range of configurations, spanning cutoff values up to $10^{-12}$ and significand lengths down to 31 bits.
The smallest achievable errors are, however, orders of magnitude larger than the FP64 machine epsilon ($\sim1.11 \times 10^{-16}$). We will elaborate on this observation in the following section when investigating the sensitivity of the main computational motifs to precision loss.

Overall, based on the targeted relative error of $10^{-3}$ for QT simulations, as previously detailed in Section \ref{sec:accuracy}, it appears feasible to run sufficiently accurate device simulations with data represented with no more than 15 bits in the significand, while simultaneously imposing cutoff values of up to $10^{-9}$. Concerning the storage of DFT+NEGF+scGW data, these findings suggest potential memory savings of up to 62.5\% when moving from FP64 to a custom FP24 format.

\section{Floating-point precision for computations}
\label{sec:compute}

In the previous section, we focused our analysis on lower-precision data formats with the goal of potentially decreasing the application's memory footprint. For their part, lower-precision computations could further reduce the time-to-solution of the application, especially after introducing lower-precision formats for storing all main quantities.
As has been established in Ref.~\cite{qtx-2025}, the computational complexity of DFT+NEGF+scGW schemes is dominated by complex matrix-matrix multiplications (ZGEMM) in the quadratic system problems of Eq.~(\ref{eq:system-solve}) for both $\mathbf{G}$ and $\mathbf{W}$, as well as in the assembly of the system matrix $\mathbf{M}(E)$ for $\mathbf{W}$ (see Table~\ref{tab:quantities}). In this section, we will therefore prioritize this basic linear algebra kernel. It should be noted that these three, ZGEMM-dominated phases
represent three distinct numerical problems, each with separate specifications and precision requirements. Throughout this part of our investigation, the DFT+NEGF+scGW data are always stored in high precision.

\subsection{Methods}
\label{sec:compute-methods}

To determine the ZGEMM precision requirements for the aforementioned three key computational motifs, we perform the following study: We simulate the same benchmark devices as in Section~\ref{sec:memory}, while mimicking lower-precision ZGEMM by truncating a number of LSBs from the significands of the operands, $\mathbf{A}$ and $\mathbf{B}$, that enter the kernel $\mathbf{C} =\mathbf{A}\mathbf{B}$.
As the actual ZGEMM are still performed in double-precision, our approach does not exactly reproduce the behavior of a ``true'' lower-precision ZGEMM kernel, but there is no hardware support for the intermediate precisions we are targeting. 
Consequently, any hardware rounding and accumulation errors that would occur in individual lower-precision floating-point operations are not accounted for in this scheme, which, nonetheless, provides useful information about precision requirements.

\begin{figure}[h] 
  \centering
  \includegraphics[width=0.48\textwidth]{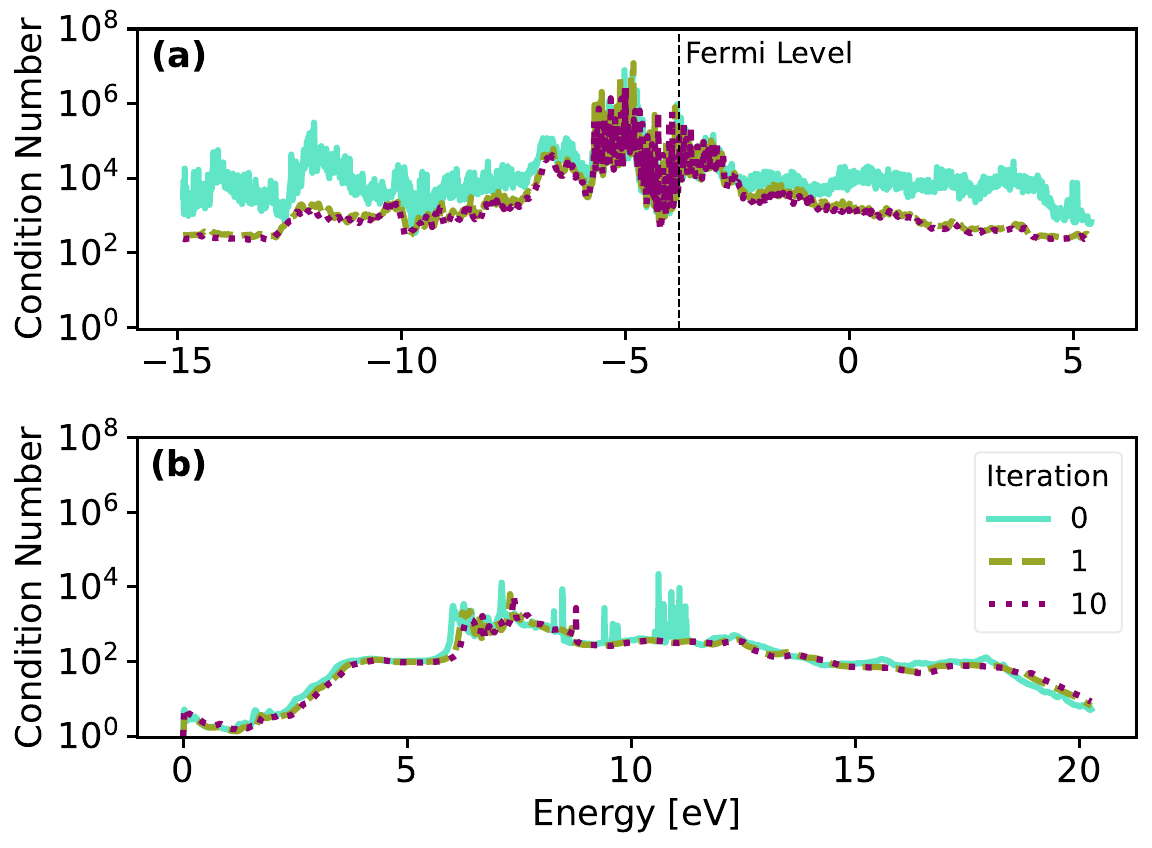}
  \caption{ 
  \normalsize
    Condition numbers of (a) the Green's function system $\mathbf{M}(E)=\left(E\mathbf{S}_{DFT} - \mathbf{H}_{DFT} - \mathbf{\Sigma}^R\right)$ and (b) the screened Coulomb interaction system $\mathbf{M}(E)=\left(\mathbf{I} -\mathbf{V}\mathbf{P}^{R}(E)\right)$ for the first (solid line), second (dashed line), and tenth (dotted line) iteration of the DFT+NEGF+scGW simulation of the \mbox{NR-S} structure. These condition numbers are shown for the energy range where they are the largest. After a few iterations, the condition numbers remain approximately constant.}
  \label{fig:condition-number}
  \vspace{-1.5em}
\end{figure}

Before starting an in-depth analysis, we can already establish preliminary error bounds by considering the numerical properties of the computational motifs we are considering. To solve Eq.~(\ref{eq:system-solve}) for $\mathbf{G}$ and $\mathbf{W}$ we rely on a form of block Gaussian elimination~\cite{maillou2026parallelquadraticselectedinversion}, where the system's condition number $\kappa(\mathbf{M})$ dictates stability and accuracy \cite{Demmel1995}.
Generally, the error bound for stabilized Gaussian elimination obeys
\begin{equation}
    \text{error} \sim \kappa(\mathbf{M}) \varepsilon_m,
    \label{eq:error}
\end{equation}
where $\varepsilon_m$ is the machine epsilon \cite{golub2013matrix}. Given the interdependencies between intermediate variables and the QT scheme's iterative nature, we can expect that this error bound is linked to the overall achievable accuracy.

In Fig.~\ref{fig:condition-number} we show the energy-resolved condition number of the system matrices $\mathbf{M}(E)$ in Eq.~(\ref{eq:system-solve}) for both $\mathbf{G}$ and $\mathbf{W}$. They are given for three iterations of the self-consistent DFT+NEGF+scGW scheme when running simulations of the \mbox{NR-S} structure as a representative example. 
We observe that $\mathbf{M}(E)$ for the first iteration of $\mathbf{G}$ is badly conditioned, with condition numbers above $10^3$ for almost all energies and up to $10^7$ in some regions.
In subsequent iterations, the condition numbers improve for most energies, but they remain elevated in an energy range close to the structure's equilibrium Fermi level $E_F$=$-3.79\;\text{eV}$.
These large condition numbers around $E_F$ could prove especially problematic when attempting to apply lower-precision computation schemes, i.e., increasing $\varepsilon_m$, as the energy range around the system's equilibrium Fermi level almost entirely determines the physical quantities of interest (electron/hole densities and electronic current).

The condition numbers for the $\mathbf{W}$ system, on the other hand, barely exceed $10^3$ over the entire energy range. While the potential for applying lower-precision computing schemes may be somewhat limited for $\mathbf{G}$, due to the high condition numbers this quantity exhibits, the quadratic system problem for $\mathbf{W}$ presents ample opportunities for optimization.

\subsection{Compute precision analysis}
\label{sec:compute-studies}

\begin{figure*}[t] 
  \centering
  \includegraphics[width=1\textwidth]{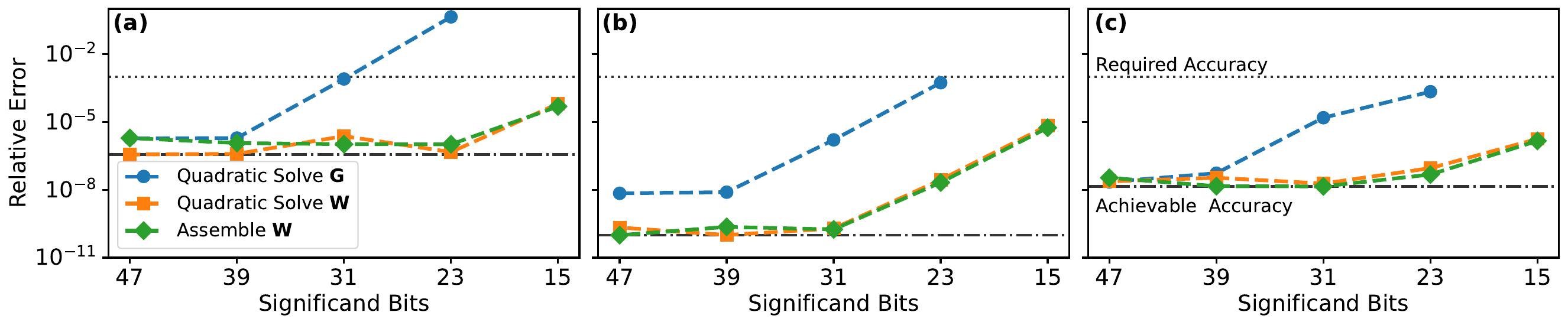}
  \caption{\normalsize
  Precision study of the quadratic solve ($\mathbf{G}$ and $\mathbf{W}$) and assembly ($\mathbf{W}$) steps of DFT+NEGF+scGW for the (a) CNT, (b) MoS$_2$, and (c) \mbox{NR-S} structures from Fig.~\ref{fig:devices}. To mimic lower-precision compute, the least significant bits are truncated in all ZGEMM inputs. The remaining significand bits are shown on the $x$-axis and the resulting error in the current on the $y$-axis.}
  \label{fig:compute_study}
  \vspace{-1.5em}
\end{figure*}

In Fig.~\ref{fig:compute_study} we present the relative error in the converged electronic current of the CNT, MoS$_2$, and \mbox{NR-S} structures with respect to the FP64 baseline as a function of the number of significand bits used to mimic lower-precision ZGEMM in the quadratic solve of $\mathbf{G}$ and $\mathbf{W}$, as well as for the system assembly of $\mathbf{W}$.
Similarly to our findings concerning lower-precision storage in Section~\ref{sec:memory-studies}, a minimal achievable error floor can be identified when lowering computational precision.
Again, these floors are specific to each tested device, and they are perfectly in line with the values found in Section~\ref{sec:memory-studies}: $10^{-6}$ for CNT, $10^{-10}$ for MoS$_2$, and $10^{-8}$ for \mbox{NR-S}.

While the relative errors in the electronic current are almost identical for all three computational motifs when using 47-bit or 39-bit significands in the CNT and \mbox{NR-S} test cases, the situation is different for MoS$_2$: Reducing the precision in the quadratic solve of $\mathbf{G}$ for this structure leads to one order-of-magnitude larger errors than the other two steps.
Moreover, when fewer than 39 bits are used to encode the significand, the relative errors in the quadratic solve of $\mathbf{G}$ rapidly increase. When attempting to truncate the significand to fewer than 23 bits in this phase of the computation, all simulations diverge.
In contrast, the relative errors for the assembly and solution of $\mathbf{W}$ are much smaller and remain almost constant for 47, 39, and 31 significand bits, for all test structures. 

Overall, we can identify a clear correlation between the condition numbers reported in Fig.~\ref{fig:condition-number} and the respective computational motif's sensitivity to lower-precision arithmetic. As already noted in Section~\ref{sec:memory-studies}, the smallest achievable errors are orders of magnitude larger than the FP64 machine epsilon, which we also attribute to the ill-conditioned numerical systems that occur in QT simulations. Furthermore, the increase of relative errors in the converged electronic current when truncating more LSBs from the significand is consistent with the error bound identified in Eq.~(\ref{eq:error}).

\vspace{-0.2em}
\section{Large Structure Evaluation}
\label{sec:evaluation}

In the previous sections, we identified the precision requirements in both data storage and key computational kernels for the DFT+NEGF+scGW method as implemented in the \textit{quatrex} application, by simulating three small, but representative benchmark structures. We want now to apply these insights to a larger, more realistic device geometry which we label \mbox{NR-L}. It is similar to \mbox{NR-S}, but is about $6\times$ longer with $4\times$ greater block sizes required for the BT tiling (3408 vs. 852 for \mbox{NR-S}). This gives rise to larger individual matrix-matrix multiplications when solving Eq.~(\ref{eq:system-solve}).
\vspace{-0.2em}
\subsection{Methods}
\label{sec:evaluation-method}

We first analyze the storage and compute requirements in isolation, but instead of merely mimicking lower precision, as in previous sections, we use the following two approaches.
First, we store the DFT+NEGF+scGW data in custom floating-point formats FPX, where $\text{X} \in [56, 48, 40, 32, 24, 16]$ refers to the total number of bits used. All stored quantities have the same number of exponent bits and bias as ``true'' FP32, while exhibiting either a reduced or increased number of significand bits, as in Section~\ref{sec:compute}.
All formats are byte-aligned to facilitate efficient integration into \textit{quatrex}, using custom compression and decompression CUDA kernels, implementing IEEE-754 rounding rules. Hence, our custom FP32 format matches IEEE-754 FP32 bitwise.
Second, since there is no hardware support for intermediate precisions between FP64 and FP32, we rely on emulation via Ozaki scheme II to achieve high-performance lower-precision matrix-matrix multiplications. We employ an implementation based on INT8 tensor cores for complex GEMM with the 3M formulation, which adds performance benefits as described in \cite{uchino2025emulation}.
In this scheme, the variable precision can be controlled by varying the number of slices $p$. Higher $p$ leads to higher precision but also higher computational cost. We sweep $p$ from 16 to 4, where 16 slices correspond ``almost'' to FP64 precision, and 4 slices are at the edge of the acceptable error tolerance of our application. Note that Ozaki scheme's accuracy depends on the data distribution, so a direct mapping between desired precision and number of slices is not straightforward. 

We separately study the impact of each approach on the results' accuracy by running simulations for up to $N_{iter}$=50 iterations and comparing the computed electronic current at each iteration to the reference FP64 execution. After 50 iterations, the DFT+NEGF+scGW method has not fully converged for the \mbox{NR-L} structure, but these trial runs allow us to infer parameters for our lower-precision scheme which we will employ in a final, large-scale simulation.

\subsection{Analysis}
\label{sec:evaluation-Analysis}

\begin{figure}[t] 
  \centering
  \includegraphics[width=0.48\textwidth]{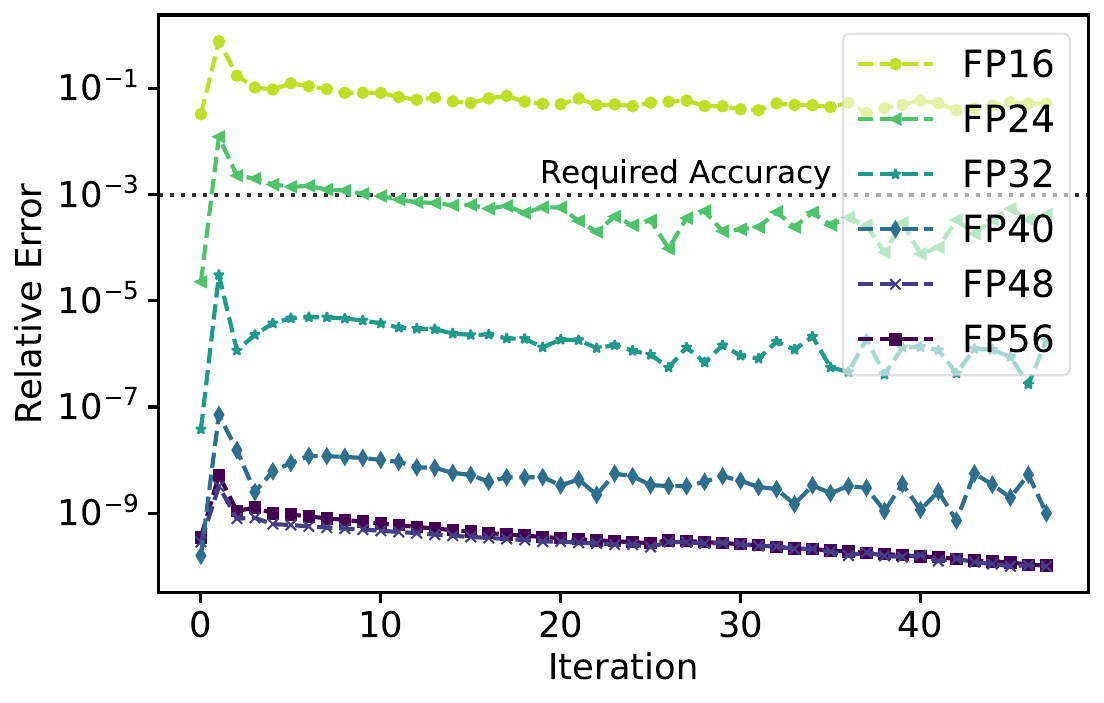}
  \caption{ 
  \normalsize
    Relative error in the electronic current flowing through the \mbox{NR-L} device with respect to the FP64 baseline as a function of the $\mathbf{G}\rightarrow\mathbf{P}\rightarrow\mathbf{W}\rightarrow\mathbf{\Sigma}$ iteration number using custom floating-point formats. The first 50 iterations are reported. Each line corresponds to having all DFT+NEGF+scGW matrices stored in FPX, where the exponent is the same as the FP32 one (8 bits), and the significand has $\text{X} - 9$ bits.}
  \label{fig:cutom_formats_nr10}
  \vspace{-1.5em}
\end{figure}

\begin{figure*}[t] 
  \centering
  \includegraphics[width=1\textwidth]{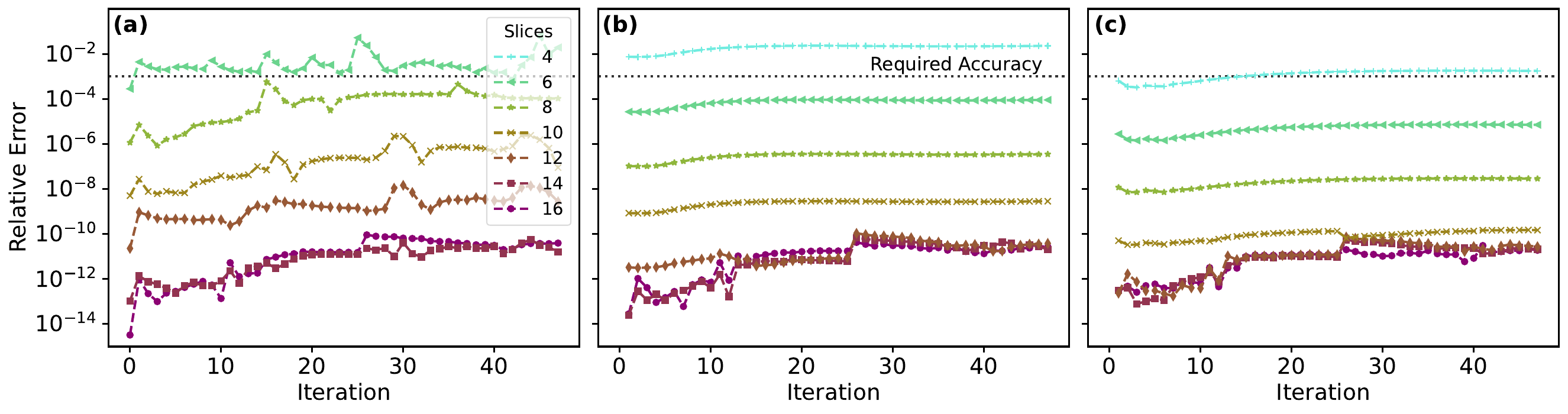}
  \vspace{-1.5em}
  \caption{ 
  \normalsize
  Same as Fig.~\ref{fig:cutom_formats_nr10}, but all variables are stored in FP64, and all ZGEMM operations are performed with the Ozaki scheme~II. (a) Quadratic solve for $\mathbf{G}$. (b) Quadratic solve for $\mathbf{W}$. (c) Assembly of the system matrix for $\mathbf{W}$. Each line corresponds to a different number of slices, with more slices providing higher accuracy.
    }
  \label{fig:ozaki-study}
  \vspace{-1.5em}
\end{figure*}

In Fig.~\ref{fig:cutom_formats_nr10}, we present the relative error in electronic current using the aforementioned FPX custom data formats. The reference is an FP64 simulation running the DFT+NEGF+scGW method for 50 self-consistent iterations.
As for the smaller devices, the error exceeds $10^{-9}$ already in the first iteration, even for the highest employed precision, FP56.
The relative error appears to peak in the second iteration before slowly decreasing thereafter.
We further notice that both FP56 and FP48 result in approximately the same errors, while the other formats are distinct.
All tested formats lead to an acceptable final accuracy, except for FP16, where the error exceeds the required accuracy of QT simulations.

Figure~\ref{fig:ozaki-study} shows relative errors in the electronic current when all ZGEMM operations during the solution of Eq.~(\ref{eq:system-solve}) and the construction of the system matrix $\mathbf{M}(E)$ for $\mathbf{W}$ are performed using Ozaki scheme~II emulation with different numbers of slices $p$. To disentangle the accuracy requirements of the individual computational motifs, in each subplot we target one of them in isolation, while the two others are treated in FP64.

We first note that the relative errors appear to increase with subsequent iterations, implying error accumulation. Still, this increase is relatively slow and shows signs of stabilization after around 30 iterations.
Furthermore, we again find that the quadratic system solve of $\mathbf{G}$ is much more sensitive to lower-precision schemes than the other operations examined, becoming unstable with fewer than 6 slices in the Ozaki scheme~II ZGEMM emulation. 
We establish that for the quadratic solve of $\mathbf{G}$, $p\leq8$ leads to results with sufficient accuracy, while for both motifs related to $\mathbf{W}$ we can employ $p=6$ while still meeting the required accuracy.

\section{Performance Results}
\label{sec:performance}

As last step, we combine the storage and performance improvement strategies that have only been considered in isolation so far. The goal of this final study is to minimize the required amount of supercomputing resources by lowering the application's memory footprint with a custom FPX format, while simultaneously increasing its computational efficiency through Ozaki scheme~II ZGEMM emulation. Practically, the findings from Figs.~\ref{fig:cutom_formats_nr10} and \ref{fig:ozaki-study} are leveraged jointly. Using the parameters determined there, we run full simulations of the \mbox{NR-L} until current conservation within 1\% is achieved, which requires approximately $N_{iter}$=440 self-consistent DFT+NEGF+scGW iterations.

\subsection{Hardware setup}
\label{sec:performance-setup}

All micro-benchmarks are run on two NVIDIA-based platforms with either GH200 (single-node) or B200 (single-gpu) GPUs, while the full simulation results are performed only on the GH200 platform since we do not have access to B200 supercomputing resources. The GH200 superchips are part of the Alps supercomputer, which is made of HPE Cray EX254n blades, each housing two nodes. Each node comprises four GH200, which combine a Hopper GPU with a 72-core Grace CPU. Each GH200 has 96\;GiB of HBM3 memory with up to 4\;TB/s bandwidth. The nodes are connected through a 25\;GB/s bidirectional bandwidth Slingshot network, while there are NVLink interconnects with 150\;GB/s bidirectional bandwidth for intranode communication. The peak FP64 tensor-core performance of the Hopper GPU is 67\;Tflop/s, while the peak INT8 one reaches 1,979\;Top/s. To access B200 GPUs, we rely on a cloud service provided by Vast.ai~\cite{vast-ai}. Each B200 has 192\;GiB of HBM3e memory with up to 8\;TB/s bandwidth. Compared to the GH200, the FP64 tensor core performance is lower at 40\;Tflop/s, while the INT8 one is more than doubled with 4,500\;Top/s. In our micro-benchmarks, we primarily run on the GPU, and thus the host CPU's features are not relevant.

\subsection{Results}
\label{sec:performance-results}

\begin{table*}[t]
\caption{
    Measured times for the main computational motifs of DFT+NEGF+scGW on a single GH200 superchip (Alps) and B200 GPU (Vast.ai~\cite{vast-ai}) for the \mbox{NR-L} device with 9 energies per GPU. 
    The workload per iteration (Tflop), performance per iteration (Tflop/s), and speed-up over the FP64 baseline are reported.
    The median values of at least 10 measurements are reported, with a relative standard deviation of less than 1\% (GH200) and 10\% (B200). 
}
\small
\centering
\begin{tabular}{l  ccccccc || ccccccc}
\hline
\multicolumn{1}{r||}{GPU} & \multicolumn{7}{c||}{GH200} & \multicolumn{7}{c}{B200}\\
\hline
\multicolumn{1}{r||}{Compute Scheme} & FP64 & \multicolumn{6}{|c||}{Ozaki II} & FP64 & \multicolumn{6}{|c}{Ozaki II}\\
\multicolumn{1}{r||}{Slices} & -- & \multicolumn{1}{|c}{16} & 14 & 12 & 10 & 8 & 6 & -- & \multicolumn{1}{|c}{16} & 14 & 12 & 10 & 8 & 6\\

\hline
\multicolumn{1}{l||}{Times [s]:} & \multicolumn{7}{c||}{} & \multicolumn{7}{c}{}\\
\multicolumn{1}{l||}{\quad \quad $\mathbf{G}$: Quad. Solve} & 22.6 & 19.3 & 17.0 & 15.2 & 13.8 & 12.3 & 10.7 & 27.2 & 12.4 & 11.8 & 10.8 & 10.1 & 9.7 & 8.8\\
\multicolumn{1}{l||}{\quad \quad $\mathbf{W}$: Assembly} & 28.1 & 28.6 & 25.5 & 22.8 & 20.9 & 18.0 & 15.5 & 43.8 & 16.2 & 15.8 & 13.2 & 12.5 & 11.6 & 10.2\\ 
\multicolumn{1}{l||}{\quad \quad $\mathbf{W}$: Quad. Solve} & 21.8 & 18.5 & 16.5 & 15.0 & 13.6 & 12.1 & 10.6 & 27.2 & 12.3 & 11.7 & 10.7 & 10.0 & 9.5 & 8.7\\
\multicolumn{1}{l||}{\quad \quad Other} & \multicolumn{7}{c||}{21.8} & \multicolumn{7}{c}{12.9}  \\
\multicolumn{1}{l||}{Total Time [s]} & 94.3 & 87.8 & 81.1 & 75.0 & 70.2 & 64.3 & 58.3  & 110.9 & 53.8 & 51.9 & 47.6 & 45.4 & 44.4 & 40.8\\
\hline
\multicolumn{1}{l||}{\makecell[l]{Speed-up\\over FP64}} & 1.0 & 1.1 & 1.2 & 1.3 & 1.3 & 1.5 & 1.6 & 1.0 & 2.1 & 2.1 & 2.3 & 2.4 & 2.5 & 2.7\\
\hline
\hline
\multicolumn{1}{l||}{Workload [Tflop]} & \multicolumn{14}{c}{3729}\\
\hline
\hline
\multicolumn{1}{l||}{Performance [Tflop/s]} & 39.5 & 42.5 & 45.9 & 49.7 & 53.1 & 58.0 & 63.9 & 33.6 & 69.3 & 71.8 & 78.4 & 82.2 & 84.0 & 91.5\\
\hline
\vspace{-2em}
\label{tab:tflop_time_flops_nr10}
\end{tabular}
\end{table*}

Table~\ref{tab:tflop_time_flops_nr10} presents the performance results for the Ozaki scheme~II emulation we implemented in \textit{quatrex}.
A performance gain over the FP64 baseline manifests for all numbers of slices $p$. The resulting speed-ups range from 1.1 to 1.6 for the GH200, and from 2.1 to 2.7 for the B200.
As expected, the B200’s speed-ups are much larger than the GH200 ones, as the FP64-to-INT8 throughput gap considerably increased between both GPU generations. 
It is worth noting that the three main compute tasks achieve higher speed-ups than the full application, as other, non-accelerated kernels consume up to 37\% of iteration time.
Also on $p$=6 slices, the performance per GPU is either close (63.9 vs. 67 Tflop/s for GH200) to or exceeds (91.5 vs. 40 Tflop/s for B200) that of the FP64 peak.

\begin{table}[h]
\small
\centering
\caption{
    Full measurement results for the \mbox{NR-L} structure combining all proposed lower-precision approaches (FPX + Ozaki scheme II) into the \textit{quatrex} tool on the Alps (GH200) supercomputer.
    Different custom data formats and slice counts are compared. The total number of energies, number of nodes (4$\times$ more GPUs), relative error in the electronic current after $N_{iter}$=380 iterations, time, workload per GPU, performance per GPU, and speed-ups with respect to FP64 are listed. If applicable, median values of at least 380 measurements are reported, with a relative standard deviation of less than 6\%.
}
\begin{tabular}{ l|| c c c c}
\hline
Data Precision  & FP64 & FP48 & FP32 & FP24 \\
\hline
Compute Scheme & FP64 & \multicolumn{3}{c}{Ozaki II}\\
\ Slices:\\
\; \makecell[l]{$\mathbf{G}$: Quad. Solve}  & -- & 14 & 10 & 8 \\
\; \makecell[l]{$\mathbf{W}$: Assembly} & -- & 12 & 8 & 6\\
\; \makecell[l]{$\mathbf{W}$: Quad. Solve}  & -- & 12 & 8 & 6\\
\hline
Energies  & \multicolumn{4}{c}{4,608} \\
Nodes & 128 & 105 & 83 & 77 \\
\hline
\makecell[l]{Energies\\per GPU} & 9 & 11 & 14 & 15 \\
\hline
\makecell[l]{Relative Error} & -- & $4.83\mathrm{e}{-10}$ & $4.01\mathrm{e}{-7}$ & $1.22\mathrm{e}{-4}$ \\
\hline
\hline
\makecell[l]{Time [s]} & 96.2 & 126.2 & 110.8 & 106.3 \\
\hline
\hline
\makecell[l]{Workload $[$Tflop$]$} & 3729 & 4557 & 5799 & 6214\\
\hline
\hline
\makecell[l]{Performance\\$\text{[Tflop/s]}$} & 38.8 & 36.1 & 52.4 & 58.4\\
\hline
\makecell[l]{Performance\\gain} & 1 & 0.93 & 1.35 & 1.51\\
\hline


\end{tabular}
\label{tab:final_runs}
\vspace{-2em}
\end{table}

In Table~\ref{tab:final_runs}, the accuracy and performance results for a full simulation of the \mbox{NR-L} device are shown, including at least 380 iterations between $\mathbf{G}$, $\mathbf{P}$, $\mathbf{W}$, and $\mathbf{\Sigma}$.
The FP64 baseline is compared to three different lower-precision schemes. The number of energy points is the same in all cases, $N_E$=4,608. This value is chosen such that the FP64 case fits into 128 nodes of Alps (512 GPUs). Thanks to the memory footprint reduction provided by the FPX formats, the number of energy points can be increased from 9 in FP64 to 15 in FP24, i.e., by a factor of 1.67$\times$, which is lower than the theoretical limit of 64/24=2.66$\times$. We attribute this shortcoming to other static memory consumptions in \textit{quatrex}.
Crucially, all lower-precision schemes achieve an accuracy below the $10^{-3}$ target and ensure convergence of the DFT+NEGF+scGW method.

The performance per GPU increases from 38.8 (FP64) to 58.4 (FP24) Tflop/s, lower than expected from the micro-benchmarks.
We attribute this degradation to lower-than-expected communication performance and greater variance when using FPX.
Still, when going from FP64 to FP24, the required computational resources can be decreased from 128 down to 77 nodes (lower memory footprint per energy), whereas the walltime only slightly increases, from 96.2 up to 106.3 sec per self-consistent iteration.
The reduction in computational power is almost entirely compensated for by the Ozaki scheme~II acceleration of matrix-matrix multiplications.
Ultimately, the speed-up per energy point is the metric that matters the most: Combining FP24 format and Ozaki scheme~II allows for a 50\% efficiency gain (speedup of 1.51$\times$).

\begin{figure}[t] 
  \centering
  \includegraphics[width=0.48\textwidth]{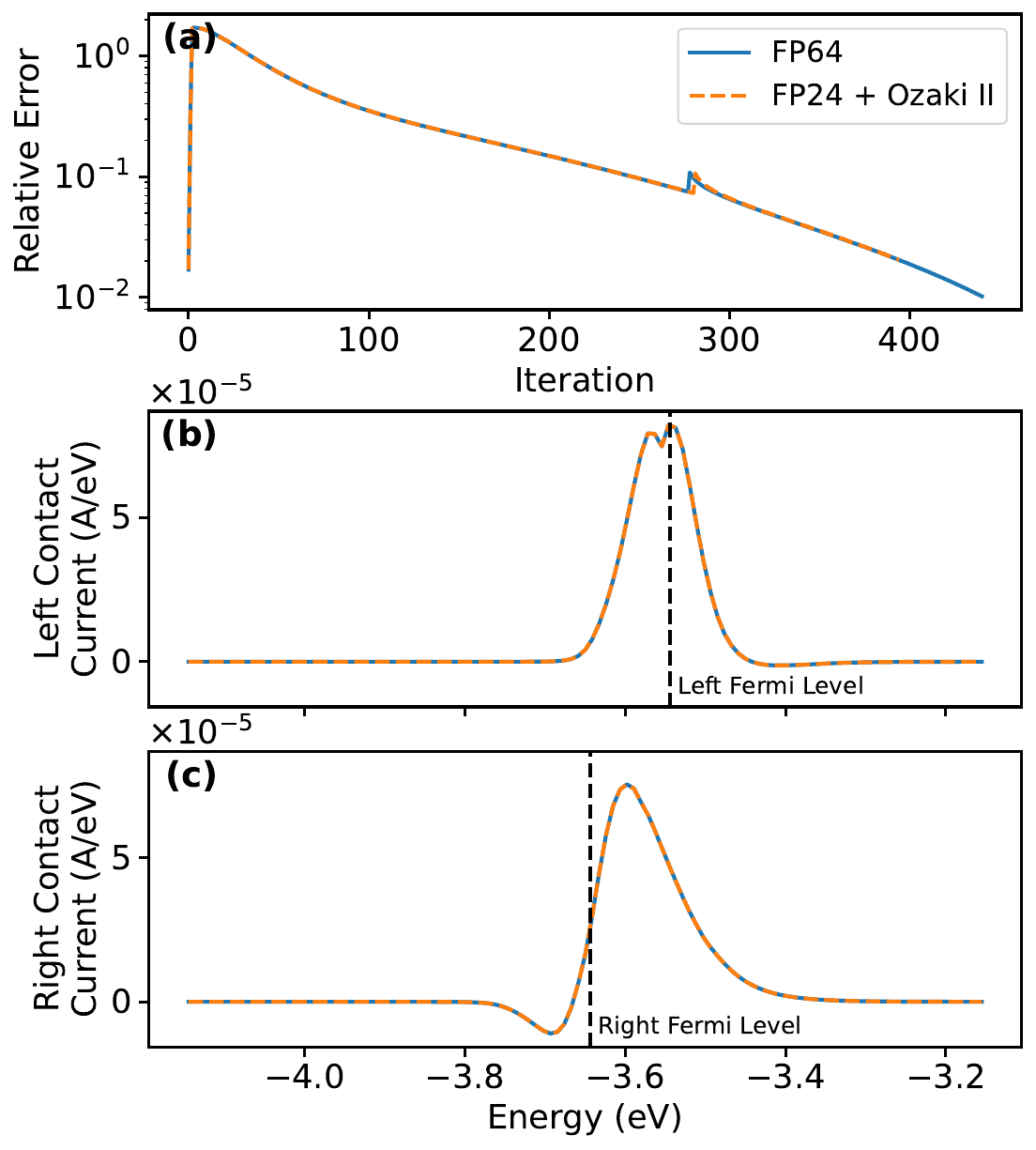}
  \caption{ 
  \normalsize
  (a) Convergence of the electronic current with respect to the self-consistent iteration number. The relative error between the left and right currents is plotted for the FP64 baseline and for the lowest-precision scheme of Table~\ref{tab:final_runs} (FP24). (b) Energy-resolved distribution of the electronic current $I_d(E)$ in the left contact for the FP64 baseline and for the FP24+Ozaki scheme~II approach. The dashed vertical line indicates the position of the left Fermi level. (c) Same as (b), but for the energy-resolved current in the right contact. The vertical dashed line corresponds to the right Fermi level.
    }
  \label{fig:current}
  \vspace{-1.5em}
\end{figure}

Importantly, the FP24+Ozaki scheme~II approach ensures satisfactory accuracy and rapid convergence, as demonstrated in Fig.~\ref{fig:current}, which plots the evolution of the left and right electronic current difference as a function of the iteration number (subplot (a)) as well as the energy-resolved electronic current $I_d(E)$ inside the left (subplot (b)) and right (subplot (c)) contacts, as obtained with this lower-precision approach and with the FP64 baseline model. We want to highlight a few remarkable points. First, subplot (a) shows that the convergence rate of the electronic current remains the same at high and low precision. Second, because of the presence of carrier-carrier interactions, the distribution of the current over the energy spectrum differs between the two contacts (subplots (b) and (c)). Third, the FP24 and FP64 curves differ by a relative error smaller than $3\cdot10^{-4}$, well within the targeted tolerance. Finally, when integrated over energy, both $I_d(E)$ lead to the same value $I_d$=8.25\;$\mu A$, demonstrating that current conservation is ensured in both the FP24 and FP64 cases. 

\section{Limitations}
\label{sec:limitations}

In this work, we conducted a comprehensive study of the numerical properties of the DFT+NEGF+scGW method across four different devices.
Still, there are limitations that we discuss below.
First, the study is primarily empirical.
Although we observe experimental evidence suggesting that the gained insights from specific, supposedly representative benchmark devices may generalize to the physical model as a whole, we cannot yet state that any specific lower-precision schemes provably work.
More importantly, we currently have no guarantees that our FP24+Ozaki scheme~II approach applies equally well to other, untested yet, structures.
Therefore, as a next step, a more theoretical investigation will be necessary, based, for example, on forward and backward error analysis, which may allow inference of the required accuracies from the input data.
However, we note that the empirical study presented here is a significant first step, demonstrating that lower precision indeed yields practical benefits.

Second, instead of our cutoff-truncation ``compression'' scheme, we could have studied other avenues, e.g., those based on tensor decompositions and low-rank approximations, or established lossy compression algorithms and libraries (such as zfp~\cite{zfp}).
Such approaches could potentially reveal further opportunities for data size reduction by, for example, grouping numbers with similar exponents, among other optimizations, which have not been explored in this work.
Here, we faced certain constraints from an implementation perspective.
Specifically, in \textit{quatrex}, the intermediate data quantities discussed are transposed through the network, and any compression solution that does not yield buffers with fine-grained access would require a significant redesign of the communication scheme.
Ultimately, we opted to proceed with a simple approach and plan the study of more advanced techniques in future work.

Third, our practical improvements, although tangible, have not yet reached the theoretical limits.
We implemented this on a best-effort basis, but performance gains at scale could be improved.
For instance, we observed poor communication performance in many cases, likely due to the use of the byte datatype for all FPX, compared to double-complex for FP64.
We note that the main focus of the work is the study of the numerical properties.
Of course, demonstrating practical benefits is also very important, which we believe is established, with further optimization potential.

\section{Related work}
\label{sec:related}

Few prior QT studies have used mixed-precision computations to some extent, specifically the OMEN QT solver~\cite{omen}, which supports models based on NEGF but also on the wave function (WF) formalism.
The WF approach formulates the QT problem as a linear system $\mathbf{A}x=b$, with only a few right-hand sides, instead of a quadratic matrix equation, as here. As such, it is limited to ballistic transport and cannot account for scattering effects, as considered here.
An early work~\cite{omen-2011} integrated a specialized parallel block-cyclic-reduction algorithm to solve the WF system, demonstrating the mixed use of double- and single-precision GEMM operations while achieving a relative error in the converged current below $10^{-2}$.
The implementation resulted in a $1.13\times$ performance increase from 1.28 Pflop/s in double-precision to 1.44 Pflop/s in mixed-precision on the Jaguar supercomputer at ORNL.
A later work~\cite{omen-2019a,omen-2019b} focused on a DFT+NEGF model capable of capturing electron-phonon interactions and of shedding light on the self-heating properties of semiconductor devices.
These interactions are computed through a convolution of the electron and phonon Green's functions, similarly to Eq.~\ref{eq:interaction}.
The work compared a double-precision convolution implementation with a half-precision one.
Taking advantage of value normalization, the relative error in the electronic current computation could be maintained below $10^{-6}$ ($10^{-3}$ without normalization).
However, since these interactions accounted for only a small fraction of the model's total runtime, the overall solver performance with the mixed-precision scheme increased by only $1.06\times$, from 86.26 Pflop/s in double-precision to 91.68 Pflop/s on the Summit supercomputer at ORNL.
Here, we note that our work improves upon these prior efforts by quantifying the required accuracy of key physical quantities to achieve different levels of precision in the simulation results.
Furthermore, our work also allows for a significant reduction in the memory requirements, enabling larger structure dimensions with finer resolution.

As far as floating-point emulation is concerned, NVIDIA provides a highly efficient GEMM implementation based on the Ozaki Scheme I~\cite{ozaki-1} since CUDA 13~\cite{cuda-cublas13}.
The performance of this routine has been showcased in NVIDIA's promotional materials~\cite{nvidia-fp-emu-blog,nvidia-fp-emu-slides} for two HPC applications, BerkeleyGW~\cite{berkeleygw} and Quantum Espresso~\cite{qe}, both from fields related to this work.
BerkleyGW is a many-body perturbation theory code to compute the excited states of solids with the GW method.
NVIDIA reported a $2.1\times$ speedup in the execution of BerkeleyGW's Epsilon module on an NVIDIA GB200-based cluster with FP64 emulation.
Quantum Espresso is an electronic-structure and materials modeling code based on DFT.
Through floating-point emulation, Quantum Espresso's execution on the Ausurf benchmark could be sped up by up to $2.98\times$ on an NVIDIA RTX PRO 6000 Blackwell Server Edition.
Our work confirms the potential of floating-point emulation, as we similarly achieve an up to $1.51\times$ speedup.

In a more general HPC context, many applications have successfully leveraged lower precision to reduce memory footprint and accelerate execution.
For example, Multi-Component Flow Code (MFC)~\cite{wilfong} employs information-geometric regularization, a method for controlling shock singularities in fluid dynamics.
Due to the method's well-conditioned numerics, a mixed-precision scheme could be developed, computing in FP32 and storing intermediates in FP16 without significant accuracy loss. This allowed one to scale many-engine spacecraft simulations, $20\times$ the size of the previous state of the art, to the full extent of the El Capitan and Alps supercomputers.
Another work on genome-wide association studies (GWAS)~\cite{ltaief} exploited the multi-precision nature of GWAS datasets to develop mixed-precision Cholesky factorization and triangular solvers.
These implementations enabled lower-precision tensor-core use, including FP8 and INT8, and led to a performance of 1.8Eop/s on almost the full Alps supercomputer.

\section{Conclusions}
\label{sec:conclusions}

We conducted a study on the numerical properties of a QT model based on the DFT+NEGF+scGW method, implemented in the open-source code \textit{quatrex}.
Through this largely experimental investigation, we demonstrated that it is possible to use lower-precision arithmetic to both reduce the immense memory requirements of scientifically relevant simulations and accelerate the main computations, while retaining sufficient accuracy in the final result.
Combining all the presented approaches, we were able to construct variants with different precision configurations, apply them to the simulation of a realistic device structure, and show that the convergence rate and properties of the physical model remain the same, i.e., the number of self-consistent iterations does not change. Overall, we achieved a $1.67\times$ reduction on the required compute resources (number of nodes) and an improvement of up to $1.51\times$ in throughput.
Future work will include a more theoretical error-analysis approach to guide the mixed-precision configuration, thus paving the way for making high-fidelity semiconductor device simulations accessible outside large supercomputers.

\section*{Acknowledgment}
This work was supported by the Swiss National Science Foundation (SNSF) under grant $\mathrm{n^\circ}$ 209358 (QuaTrEx) and grant $\mathrm{n^\circ}$~205602 (NCCR MARVEL), and by the Platform for Advanced Scientific Computing in Switzerland (BoostQT). We acknowledge support from CSCS (projects c33, lp16, lp82).
The authors would like to especially thank Tim Robinson (CSCS) for access to and support of the computational resources.

\bibliographystyle{IEEEtran}
\bibliography{refs}

@inproceedings{omen-2011,
author = {Luisier, Mathieu and Boykin, Timothy B. and Klimeck, Gerhard and Fichtner, Wolfgang},
title = {Atomistic nanoelectronic device engineering with sustained performances up to 1.44 PFlop/s},
year = {2011},
isbn = {9781450307710},
publisher = {Association for Computing Machinery},
address = {New York, NY, USA},
url = {https://doi.org/10.1145/2063384.2063387},
doi = {10.1145/2063384.2063387},
booktitle = {Proceedings of 2011 International Conference for High Performance Computing, Networking, Storage and Analysis},
articleno = {2},
numpages = {11},
location = {Seattle, Washington},
series = {SC '11}
}

@inproceedings{omen-2019a,
author = {Ziogas, Alexandros Nikolaos and Ben-Nun, Tal and Fern\'{a}ndez, Guillermo Indalecio and Schneider, Timo and Luisier, Mathieu and Hoefler, Torsten},
title = {A data-centric approach to extreme-scale ab initio dissipative quantum transport simulations},
year = {2019},
isbn = {9781450362290},
publisher = {Association for Computing Machinery},
address = {New York, NY, USA},
url = {https://doi.org/10.1145/3295500.3357156},
doi = {10.1145/3295500.3357156},
booktitle = {Proceedings of the International Conference for High Performance Computing, Networking, Storage and Analysis},
articleno = {1},
numpages = {13},
location = {Denver, Colorado},
series = {SC '19}
}

@inproceedings{omen-2019b,
author = {Ziogas, Alexandros Nikolaos and Ben-Nun, Tal and Fern\'{a}ndez, Guillermo Indalecio and Schneider, Timo and Luisier, Mathieu and Hoefler, Torsten},
title = {Optimizing the data movement in quantum transport simulations via data-centric parallel programming},
year = {2019},
isbn = {9781450362290},
publisher = {Association for Computing Machinery},
address = {New York, NY, USA},
url = {https://doi.org/10.1145/3295500.3356200},
doi = {10.1145/3295500.3356200},
booktitle = {Proceedings of the International Conference for High Performance Computing, Networking, Storage and Analysis},
articleno = {78},
numpages = {17},
location = {Denver, Colorado},
series = {SC '19}
}

@inproceedings{qtx-2024,
author = {Deuschle, Leonard and Maeder, Alexander and Maillou, Vincent and Vetsch, Nicolas and Winka, Anders and Cao, Jiang and Ziogas, Alexandros Nikolaos and Luisier, Mathieu},
title = {Towards Exascale Simulations of Nanoelectronic Devices in the GW Approximation},
year = {2024},
isbn = {9798350352917},
publisher = {IEEE Press},
url = {https://doi.org/10.1109/SC41406.2024.00069},
doi = {10.1109/SC41406.2024.00069},
booktitle = {Proceedings of the International Conference for High Performance Computing, Networking, Storage, and Analysis},
articleno = {63},
numpages = {16},
location = {Atlanta, GA, USA},
series = {SC '24}
}

@inproceedings{qtx-2025,
author = {Vetsch, Nicolas and Maeder, Alexander and Maillou, Vincent and Winka, Anders and Cao, Jiang and Kwasniewski, Grzegorz and Deuschle, Leonard and Hoefler, Torsten and Ziogas, Alexandros Nikolaos and Luisier, Mathieu},
title = {Ab-initio Quantum Transport with the GW Approximation, 42,240 Atoms, and Sustained Exascale Performance},
year = {2025},
isbn = {9798400714665},
publisher = {Association for Computing Machinery},
address = {New York, NY, USA},
url = {https://doi.org/10.1145/3712285.3771784},
doi = {10.1145/3712285.3771784},
booktitle = {Proceedings of the International Conference for High Performance Computing, Networking, Storage and Analysis},
pages = {1–13},
numpages = {13},
location = {
},
series = {SC '25}
}

@ARTICLE{ieee-754,
  author={},
  journal={IEEE Std 754-2019 (Revision of IEEE 754-2008)}, 
  title={IEEE Standard for Floating-Point Arithmetic}, 
  year={2019},
  volume={},
  number={},
  pages={1-84},
  doi={10.1109/IEEESTD.2019.8766229}}

@article{deuschle2025electron,
  title = {Electron-electron interactions in device simulation via nonequilibrium Green's functions and the GW approximation},
  author = {Deuschle, Leonard and Cao, Jiang and Ziogas, Alexandros Nikolaos and Winka, Anders and Maeder, Alexander and Vetsch, Nicolas and Luisier, Mathieu},
  journal = {Phys. Rev. B},
  volume = {111},
  issue = {19},
  pages = {195421},
  numpages = {18},
  year = {2025},
  month = {May},
  publisher = {American Physical Society},
  doi = {10.1103/PhysRevB.111.195421},
  url = {https://doi.org/10.1103/PhysRevB.111.195421}
}

@manual{nvidia_gh200_ds,
  title        = {NVIDIA GH200 Grace Hopper Superchip Datasheet},
  author       = {{NVIDIA Corporation}},
  year         = {2025},
  url          = {https://nvdam.widen.net/s/rrgqqnpbz8/grace-datasheet-gh200-grace-hopper-superchip-3773000},
  note         = {Accessed: 2026-03-25}
}

@manual{nvidia_gb200_ds,
  title        = {NVIDIA Blackwell Datasheet},
  author       = {{NVIDIA Corporation}},
  year         = {2025},
  url          = {https://nvdam.widen.net/s/wwnsxrhm2w/blackwell-datasheet-3384703},
  note         = {Accessed: 2026-03-25}
}

@misc{top500,
  author = {Dongarra, Jack and Meuer, Hans and Strohmaier, Erich},
  title = {{TOP500}},
  year = {2025},
  howpublished = {\url{https://www.top500.org}},
  note = {Accessed: 2026-03-25}
}

@INPROCEEDINGS{intel-nrfet,
  author={Agrawal, A. and Chakraborty, W. and Li, W. and Ryu, H. and Markman, B. and Hoon, S. H. and Paul, R. K and Huang, C. Y. and Choi, S. M. and Rho, K. and Shu, A. and Iglesias, R. and Wallace, P. and Ghosh, S. and Cheong, K. L. and Hockel, J. L. and Thorman, R. and Baumgartel, L. and Shoer, L. and Mishra, V. and Berrada, S. and Ashita, A. and Weber, C. and Obradovic, B. and Oni, A. A. and Brooks, Z. and Franco, N. and Kavalieros, J. and Dewey, G.},
  booktitle={2024 IEEE International Electron Devices Meeting (IEDM)}, 
  title={Silicon RibbonFET CMOS at 6nm Gate Length}, 
  year={2024},
  doi={10.1109/IEDM50854.2024.10873367},
  url={https://doi.org/10.1109/IEDM50854.2024.10873367}}

@article{ozaki-1,
author = {Ozaki, Katsuhisa and Ogita, Takeshi and Oishi, Shin'Ichi and Rump, Siegfried M.},
title = {Error-free transformations of matrix multiplication by using fast routines of matrix multiplication and its applications},
year = {2012},
issue_date = {January   2012},
publisher = {Springer-Verlag},
address = {Berlin, Heidelberg},
volume = {59},
number = {1},
issn = {1017-1398},
url = {https://doi.org/10.1007/s11075-011-9478-1},
doi = {10.1007/s11075-011-9478-1},
journal = {Numer. Algorithms},
month = jan,
pages = {95–118},
numpages = {24}
}

@misc{ozaki-2,
      title={Ozaki Scheme II: A GEMM-oriented emulation of floating-point matrix multiplication using an integer modular technique}, 
      author={Katsuhisa Ozaki and Yuki Uchino and Toshiyuki Imamura},
      year={2025},
      eprint={2504.08009},
      archivePrefix={arXiv},
      primaryClass={cs.MS},
      url={https://doi.org/10.48550/arXiv.2504.08009},
}

@article{ootomo2024dgemm,
    author = {Hiroyuki Ootomo and Katsuhisa Ozaki and Rio Yokota},
    title = {DGEMM on integer matrix multiplication unit},
    journal = {The International Journal of High Performance Computing Applications},
    year = {2024},
    volume={38},
    number={4},
    pages={297--313},
    doi = {10.1177/10943420241239588},
    URL = {https://doi.org/10.1177/10943420241239588},
    publisher={SAGE Publications},
}

@article{uchino2025performance,
   title={Performance enhancement of the Ozaki Scheme on integer matrix multiplication unit},
   volume={39},
   ISSN={1741-2846},
   url={https://doi.org/10.1177/10943420241313064},
   DOI={10.1177/10943420241313064},
   number={3},
   journal={The International Journal of High Performance Computing Applications},
   publisher={SAGE Publications},
   author={Uchino, Yuki and Ozaki, Katsuhisa and Imamura, Toshiyuki},
   year={2025},
   month={jan},
   pages={462–476}
}

@inproceedings{schwarz2026guaranteed,
author = {Schwarz, Angelika and Anders, Anton and Brower, Cole and Bayraktar, Harun and Gunnels, John and Clark, Kate and Xu, RuQing G. and Rodriguez, Samuel and Cayrols, Sebastien and Tabaszewski, Pawel and Podlozhnyuk, Victor},
title = {Guaranteed DGEMM Accuracy While Using Reduced Precision Tensor Cores Through Extensions of the Ozaki Scheme},
year = {2026},
isbn = {9798400720673},
publisher = {Association for Computing Machinery},
address = {New York, NY, USA},
url = {https://doi.org/10.1145/3773656.3773670},
doi = {10.1145/3773656.3773670},
booktitle = {Proceedings of the Supercomputing Asia and International Conference on High Performance Computing in Asia Pacific Region},
pages = {91–101},
numpages = {11},
location = {
},
series = {SCA/HPCAsia '26}
}

@article{Hybertsen1986,
  title = {Electron correlation in semiconductors and insulators: Band gaps and quasiparticle energies},
  author = {Hybertsen, Mark S. and Louie, Steven G.},
  journal = {Phys. Rev. B},
  volume = {34},
  issue = {8},
  pages = {5390--5413},
  numpages = {0},
  year = {1986},
  month = {Oct},
  publisher = {American Physical Society},
  doi = {10.1103/PhysRevB.34.5390},
  url = {https://doi.org/10.1103/PhysRevB.34.5390}
}

@article{Thygesen2008,
  title = {Conserving $GW$ scheme for nonequilibrium quantum transport in molecular contacts},
  author = {Thygesen, Kristian S. and Rubio, Angel},
  journal = {Phys. Rev. B},
  volume = {77},
  issue = {11},
  pages = {115333},
  numpages = {22},
  year = {2008},
  month = {Mar},
  publisher = {American Physical Society},
  doi = {10.1103/PhysRevB.77.115333},
  url = {https://doi.org/10.1103/PhysRevB.77.115333}
}

@inbook{datta_1995, place={Cambridge}, series={Cambridge Studies in Semiconductor Physics and Microelectronic Engineering}, title={Non-equilibrium Green's function formalism}, DOI={10.1017/CBO9780511805776.009}, booktitle={Electronic Transport in Mesoscopic Systems}, publisher={Cambridge University Press}, author={Datta, Supriyo}, year={1995}, pages={293–342}, collection={Cambridge Studies in Semiconductor Physics and Microelectronic Engineering},
url={https://doi.org/10.1017/CBO9780511805776.009}}

@article{dft,
  title = {Self-{{Consistent Equations Including Exchange}} and {{Correlation Effects}}},
  author = {Kohn, W. and Sham, L. J.},
  year = {1965},
  month = nov,
  journal = {Phys. Rev.},
  volume = {140},
  number = {4A},
  pages = {A1133-A1138},
  publisher = {American Physical Society},
  doi = {10.1103/PhysRev.140.A1133},
  url={https://doi.org/10.1103/PhysRev.140.A1133},
  urldate = {2023-11-10}
}

@INPROCEEDINGS{mos2_iedm_2025,
  author={Barraud, S. and Rodriguez-Fano, M. and Pedini, J.M. and Cadot, S. and Chouk, R. and Dey, B. and Hartmann, J.M. and Gharbi, A. and Comboroure, C. and Sarrazin, A. and Boulard, F. and Laraignou, L. and Campo, A. and Grampeix, H. and Castan, C. and Sturm, J. and Souhaité, A. and Lassenberger, A. and Couture, L. and Mariolle, D. and Hauchecorne, P. and Loup, V. and Gapihan, E. and O’Brien, K.P. and Avci, U. and Andrieu, F.},
  booktitle={2025 IEEE International Electron Devices Meeting (IEDM)}, 
  title={Novel channel-last integration of ALD MoS2 into stacked channel FETs on 300mm wafers}, 
  year={2025},
  volume={},
  number={},
  pages={1-4},
  doi={10.1109/IEDM50572.2025.11353602},
  url={https://doi.org/10.1109/IEDM50572.2025.11353602}
}

@INPROCEEDINGS{cnt_vlsi_2023,
  author={Pitner, Gregory and Safron, Nathaniel and Chao, Tzu-Ang and Li, Shengman and Su, Sheng-Kai and Zeevi, Gilad and Lin, Qing and Chiu, Hsin-Yuan and Passlack, Matthias and Zhang, Zichen and Sathaiya, D. Mahaveer and Wei, Aslan and Gilardi, Carlo and Chen, Edward and Liew, San-Lin and Hou, Vincent D.-H. and Wu, Chung-Wei and Wu, Jeff and Lin, Zhiwei and Fagan, Jeffrey and Zheng, Ming and Wang, Han and Mitra, Subhasish and Philip Wong, H.-S. and Radu, Iuliana},
  booktitle={2023 IEEE Symposium on VLSI Technology and Circuits (VLSI Technology and Circuits)}, 
  title={Building high performance transistors on carbon nanotube channel}, 
  year={2023},
  volume={},
  number={},
  pages={1-2},
  doi={10.23919/VLSITechnologyandCir57934.2023.10185374},
  url={https://doi.org/10.23919/VLSITechnologyandCir57934.2023.10185374}
  }

@article{stokbro,
  title = {Density-functional method for nonequilibrium electron transport},
  author = {Brandbyge, Mads and Mozos, Jos\'e-Luis and Ordej\'on, Pablo and Taylor, Jeremy and Stokbro, Kurt},
  journal = {Phys. Rev. B},
  volume = {65},
  issue = {16},
  pages = {165401},
  numpages = {17},
  year = {2002},
  month = {Mar},
  publisher = {American Physical Society},
  doi = {10.1103/PhysRevB.65.165401},
  url = {https://doi.org/10.1103/PhysRevB.65.165401}
}

@article{frederiksen,
  title = {Inelastic transport theory from first principles: Methodology and application to nanoscale devices},
  author = {Frederiksen, Thomas and Paulsson, Magnus and Brandbyge, Mads and Jauho, Antti-Pekka},
  journal = {Phys. Rev. B},
  volume = {75},
  issue = {20},
  pages = {205413},
  numpages = {22},
  year = {2007},
  month = {May},
  publisher = {American Physical Society},
  doi = {10.1103/PhysRevB.75.205413},
  url = {https://doi.org/10.1103/PhysRevB.75.205413}
}

@article{Dossena2025,
  author  = {Dossena, M. and Van Troeye, B. and Ducry, F. and Cao, J. and Afzalian, A. and Pourtois, G. and Luisier, M.},
  title   = {Mobility calculation in disordered {WS}$_2$-{Al}$_2${O}$_3$ stacks from first principles},
  journal = {npj 2D Materials and Applications},
  year    = {2025},
  volume  = {9},
  number  = {1},
  pages   = {67},
  doi     = {10.1038/s41699-025-00587-9},
  url     = {https://doi.org/10.1038/s41699-025-00587-9}
}

@misc{maillou2026parallelquadraticselectedinversion,
      title={Parallel Quadratic Selected Inversion in Quantum Transport Simulation}, 
      author={Vincent Maillou and Matthias Bollhofer and Olaf Schenk and Alexandros Nikolaos Ziogas and Mathieu Luisier},
      year={2026},
      eprint={2601.04904},
      archivePrefix={arXiv},
      primaryClass={cs.DC},
      url={https://doi.org/10.48550/arXiv.2601.04904}, 
}

@book{Kadanoff2018,
  doi = {10.1201/9780429493218},
  url = {https://doi.org/10.1201/9780429493218},
  year = {2018},
  month = mar,
  publisher = {{CRC} Press},
  author = {Leo P. Kadanoff and Gordon Baym},
  title = {Quantum Statistical Mechanics}
}

@misc{cuda-cublas13,
  author = {{NVIDIA Corporation}},
  title = {{cuBLAS 13.0 documentation}},
  year = {2025},
  howpublished = {\url{https://docs.nvidia.com/cuda/archive/13.0.2/cublas/index.html#floating-point-emulation}},
  note = {Accessed: 2026-03-31}
}

@misc{nvidia-fp-emu-blog,
  author = {{NVIDIA Corporation}},
  title = {{Unlocking Tensor Core Performance with Floating Point Emulation in cuBLAS | NVIDIA Technical Blog}},
  year = {2025},
  howpublished = {\url{https://developer.nvidia.com/blog/unlocking-tensor-core-performance-with-floating-point-emulation-in-cublas}},
  note = {Accessed: 2026-03-31}
}

@misc{nvidia-fp-emu-slides,
  author = {{NVIDIA Corporation}},
  title = {{Floating Point Emulation in NVIDIA Math Libraries}},
  year = {2025},
  howpublished = {\url{https://indico.cern.ch/event/1538409/contributions/6521976/attachments/3096181/5485165/cern-talk.pdf}},
  note = {Accessed: 2026-03-31}
}

@INPROCEEDINGS{berkeleygw,
  author={Ben, Mauro Del and Yang, Charlene and Li, Zhenglu and Jornada, Felipe H. da and Louie, Steven G. and Deslippe, Jack},
  booktitle={SC20: International Conference for High Performance Computing, Networking, Storage and Analysis}, 
  title={Accelerating Large-Scale Excited-State GW Calculations on Leadership HPC Systems}, 
  year={2020},
  volume={},
  number={},
  pages={1-11},
  doi={10.1109/SC41405.2020.00008},
  url={https://doi.org/10.1109/SC41405.2020.00008}
}

@article{qe,
    author = {Giannozzi, Paolo and Baseggio, Oscar and Bonfà, Pietro and Brunato, Davide and Car, Roberto and Carnimeo, Ivan and Cavazzoni, Carlo and de Gironcoli, Stefano and Delugas, Pietro and Ferrari Ruffino, Fabrizio and Ferretti, Andrea and Marzari, Nicola and Timrov, Iurii and Urru, Andrea and Baroni, Stefano},
    title = {Quantum ESPRESSO toward the exascale},
    journal = {The Journal of Chemical Physics},
    volume = {152},
    number = {15},
    pages = {154105},
    year = {2020},
    month = {04},
    issn = {0021-9606},
    doi = {10.1063/5.0005082},
    url = {https://doi.org/10.1063/5.0005082},
    eprint = {https://pubs.aip.org/aip/jcp/article-pdf/doi/10.1063/5.0005082/20749723/154105_1_5.0005082.pdf},
}

@inproceedings{wilfong,
author = {Wilfong, Benjamin and Radhakrishnan, Anand and Le Berre, Henry and Vickers, Daniel and Prathi, Tanush and Tselepidis, Nikolaos and Dorschner, Benedikt and Budiardja, Reuben and Cornille, Brian and Abbott, Stephen and Sch\"{a}fer, Florian and Bryngelson, Spencer},
title = {Simulating many-engine spacecraft: Exceeding 1 quadrillion degrees of freedom via information geometric regularization},
year = {2025},
isbn = {9798400714665},
publisher = {Association for Computing Machinery},
address = {New York, NY, USA},
url = {https://doi.org/10.1145/3712285.3771783},
doi = {10.1145/3712285.3771783},
booktitle = {Proceedings of the International Conference for High Performance Computing, Networking, Storage and Analysis},
pages = {14–24},
numpages = {11},
location = {
},
series = {SC '25}
}

@inproceedings{ltaief,
author = {Ltaief, Hatem and Alomairy, Rabab and Cao, Qinglei and Ren, Jie and Slim, Lotfi and Kurth, Thorsten and Dorschner, Benedikt and Bougouffa, Salim and Abdelkhalak, Rached and Keyes, David E.},
title = {Toward Capturing Genetic Epistasis From Multivariate Genome-Wide Association Studies Using Mixed-Precision Kernel Ridge Regression},
year = {2024},
isbn = {9798350352917},
publisher = {IEEE Press},
url = {https://doi.org/10.1109/SC41406.2024.00012},
doi = {10.1109/SC41406.2024.00012},
booktitle = {Proceedings of the International Conference for High Performance Computing, Networking, Storage, and Analysis},
articleno = {6},
numpages = {12},
location = {Atlanta, GA, USA},
series = {SC '24}
}

@article{omen,
  title = {Atomistic simulation of nanowires in the $s{p}^{3}{d}^{5}{s}^{*}$ tight-binding formalism: From boundary conditions to strain calculations},
  author = {Luisier, Mathieu and Schenk, Andreas and Fichtner, Wolfgang and Klimeck, Gerhard},
  journal = {Phys. Rev. B},
  volume = {74},
  issue = {20},
  pages = {205323},
  numpages = {12},
  year = {2006},
  month = {Nov},
  publisher = {American Physical Society},
  doi = {10.1103/PhysRevB.74.205323},
  url = {https://doi.org/10.1103/PhysRevB.74.205323}
}

@misc{uchino2025emulation,
      title={Emulation of Complex Matrix Multiplication based on the Chinese Remainder Theorem}, 
      author={Yuki Uchino and Qianxiang Ma and Toshiyuki Imamura and Katsuhisa Ozaki and Patrick Lars Gutsche},
      year={2025},
      eprint={2512.08321},
      archivePrefix={arXiv},
      primaryClass={cs.DC},
      url={https://doi.org/10.48550/arXiv.2512.08321}, 
}

@book{golub2013matrix,
  author = {Golub, Gene H. and van Loan, Charles F.},
  edition = {Fourth},
  isbn = {1421407949 9781421407944},
  publisher = {JHU Press},
  refid = {824733531},
  title = {Matrix Computations},
  doi = {10.1137/1.9781421407944},
  url = {https://epubs.siam.org/doi/abs/10.1137/1.9781421407944},
  year = 2013
}

@article{Fasi2021,
  title = {Numerical behavior of NVIDIA tensor cores},
  volume = {7},
  ISSN = {2376-5992},
  url = {https://doi.org/10.7717/peerj-cs.330},
  DOI = {10.7717/peerj-cs.330},
  journal = {PeerJ Computer Science},
  publisher = {PeerJ},
  author = {Fasi,  Massimiliano and Higham,  Nicholas J. and Mikaitis,  Mantas and Pranesh,  Srikara},
  year = {2021},
  month = feb,
  pages = {e330}
}

@misc{vast-ai,
  author = {{Vast.ai}},
  title = {Vast.ai: Rent GPUs},
  year = {2026},
  url = {https://vast.ai/},
  note = {Accessed: April 5, 2026}
}

@article{Demmel1995,
  title = {Stability of block LU factorization},
  volume = {2},
  ISSN = {1099-1506},
  url = {https://doi.org/10.1002/nla.1680020208},
  DOI = {10.1002/nla.1680020208},
  number = {2},
  journal = {Numerical Linear Algebra with Applications},
  publisher = {Wiley},
  author = {Demmel,  James W. and Higham,  Nicholas J. and Schreiber,  Robert S.},
  year = {1995},
  month = mar,
  pages = {173–190}
}

@ARTICLE{zfp,
  author={Lindstrom, Peter},
  journal={IEEE Transactions on Visualization and Computer Graphics}, 
  title={Fixed-Rate Compressed Floating-Point Arrays}, 
  year={2014},
  volume={20},
  number={12},
  pages={2674-2683},
  doi={10.1109/TVCG.2014.2346458},
  url={https://doi.org/10.1109/TVCG.2014.2346458}
}

@article{kirchhoff,
author = {Kirchhoff, Studiosus},
title = {Ueber den Durchgang eines elektrischen Stromes durch eine Ebene, insbesondere durch eine kreisförmige},
journal = {Annalen der Physik},
volume = {140},
number = {4},
pages = {497-514},
url = {https://doi.org/10.1002/andp.18451400402},
eprint = {https://onlinelibrary.wiley.com/doi/pdf/10.1002/andp.18451400402},
year = {1845}
}

\end{document}